\documentclass[prd,aps,twocolumn,showpacs,nofootinbib]{revtex4-1}

\usepackage{graphicx}
\usepackage{epsfig}

\newcommand{\Vcb}{|\ensuremath{V_{\mathrm{cb}}}|}

\begin{document}

\title{Heavy and light meson wavefunctions}

\author{Xing-Gang Wu}
\email{wuxg@cqu.edu.cn}
\affiliation{Department of Physics, Chongqing University, Chongqing 401331, P.R. China}

\author{Tao Huang}
\email{huangtao@ihep.ac.cn}
\affiliation{Institute of High Energy Physics and Theoretical Physics Center for Science Facilities, Chinese Academy of Sciences, Beijing 100049, P.R. China}

\date{\today}

\begin{abstract}

We present a short review on the properties of heavy and light mesons' light-cone wavefunctions (LCWFs), and their distribution amplitudes (DAs). The B meson LCWFs can be treated by taking the heavy quark limit ($m_b\to\infty$) and by using the heavy quark effective theory. Furthermore, we propose a simple model for the B meson WFs with 3-particle Fock states' contributions, whose behaviors are controlled by two parameters $\bar\Lambda$ and $\delta$. Using such model, the form factors $F^{B\to\pi}_{+,0,T}$ and $F^{B\to K}_{+,0,T}$ in large recoil region are studied up to ${\cal O}(1/m_b^2)$ within the $k_T$ factorization approach. On the other hand, we adopt Brodsky-Huang-Lepage (BHL) prescription for constructing the WFs of the lighter pseudoscalars as $\eta_c$, D-meson, pion, kaon, $\eta^{(\prime)}$ and etc. Such BHL-like model can be conveniently extended to construct the LCWFs for light scalar or vector mesons. Within such model the longitudinal distributions of those WFs are basically determined by a parameter $B$, whose value can be determined via a global fit of experimental data.


\end{abstract}
\maketitle
%
%\tableofcontents
%
\section{Introduction}

The hadronic light-cone wavefunctions (LCWFs) exhibit all properties of the bound state and provide underlying links between the hadronic phenomena at large and small distances~\cite{brodsky}. They are universal physical quantities for applying pQCD to exclusive processes. The study of phenomenology of hadronic LCWFs is an essential task to understand QCD dynamics, especially to understand the QCD factorization theory. Right now, we have not enough knowledge to determine the hadron LCWFs or their distribution amplitudes (DAs). The nonperturbative models such as AdS/QCD and light-front holography may provide a solution, a very recent review of which can be found in Ref.\cite{ads}. Practically, one can study their properties under some approximations/prescriptions or via a global fit of known experimental processes.

The B meson LCWFs can be treated by taking the heavy quark limit, $m_b\to\infty$, and by using the heavy quark effective theory (HQET)~\cite{hqet1,hqet2,hqet3}. The B meson DAs and WFs have been investigated by various approaches~\cite{huangB1,qiao0,beneke,descotes,bdistribution1,braun,
bdistribution2,geyer,alex,libwave,qiao,bwave,bwave2,bwave3,bwave4,bwave5,bwave6}. Refs.\cite{qiao0,qiao,bwave,bwave2} present an analytic solution for the B meson WFs $\Psi_{\pm}(\omega,z^2)$ up to next-to-leading order in Fock state expansion, which satisfies the constraints from the QCD equations of motion and the heavy-quark symmetry~\cite{heavyquark,heavyquark2,heavyquark3}. Under the Wandzura-Wilczek (WW) approximation~\cite{ww}, which corresponds to valence quark distribution only, the B meson WFs can be determined uniquely in terms of the ``effective mass'' ($\bar{\Lambda}$) and its transverse momentum dependence behaves simply as a $\delta$-function. Such transverse momentum distribution can be considerably broadened by including the 3-particle Fock states' contributions. In the present paper, we shall review the properties of B meson WFs/DAs derived under HQET, whose parameters can be further constrained through a comparative study of $B\to\pi/K$ transition form factors (TFFs) derived under various approaches, such as pQCD, QCD light-cone sum rules (LCSRs) and lattice QCD, or by a comparison of theoretical estimations with the experimental data on B decays.

For the light mesons as $\pi$, $K$, $\eta^{(\prime)}$ and etc., one can adopt the Brodsky-Huang-Lepage (BHL) prescription~\cite{bhl1,bhl2,bhl3} together with the Wigner-Melosh rotation effect~\cite{wigner} to construct a reliable model for their WFs. This is the so-called revised light-cone harmonic oscillator model. The BHL prescription is obtained in a way by connecting the equal-time WF in the rest frame and the WF in the infinite momentum frame. Because of Wigner-Melosh rotation, there are higher-order helicity components for the WFs. Even though these higher helicity components are usually power suppressed in large energy region, they shall lead to sizable contributions in low and intermediate energy regions. It has been observed that the introduction of higher helicity components into the light-cone formalism shall result in significant consequence in several problems concerning the applicability of perturbative QCD, such as the pion and kaon electro-magnetic form factors~\cite{pi1,pi2,pi3,pi4,kaon1}, the pion-photon TFF~\cite{TFF1,TFF2,TFF3,TFF4,TFF5,TFF6,TFF7,TFF8,TFF9,TFF10}, and etc.. It has been found that by a proper change of input parameters, especially the value of $B$ that basically determines the longitudinal behavior of the light meson WFs, one can conveniently simulate the shape of the light meson's DA from asymptotic-like~\cite{brodsky} to CZ-like~\cite{cz}. Thus, it provides a convenient way to compare the estimates with different DA behaviors. That is, by comparing the theoretical estimations with the corresponding experimental data, those undetermined parameters and hence its DA behavior can be fixed or at least be greatly restricted.

For the lighter mesons involving charm quark as compared to the B meson, i.e. the charmonium $\eta_c$ or $J/\psi$ and the D meson, the conditions are quite different from the case of B meson, since the charm quark is not heavy enough~\cite{charm1,charm2,charm3,charm4,charm5,charm6,bwave4,charm8,charm9,charm10,charm11,charm12}. We can construct their WFs in a similar way as that of light mesons. For example, it has been shown \cite{charm12} that one can obtain a reliable LCSRs estimation for exclusive process $e^{+}+e^{-}\rightarrow J/\psi+\eta_c$ by using a realistic $\eta_c$ DA. That is, a proper $\eta_c$ DA at a certain energy scale can result in a compatible prediction with the Belle and BaBar experimental data~\cite{charm5,charm12}. Furthermore, it is noted that a proper way of constructing the D meson WF with a better end-point behavior at small longitudinal and transverse distribution region is very important for dealing with high energy processes such as $D\to\pi l\nu$ and $B\to D l\nu$. Especially, the $B\to D l\nu$ process can be further used to derive the value of $\Vcb$~\cite{expV1,expV2,expV3}. In the present paper, we shall discuss the $\eta_c$ and D-meson WFs/DAs in detail, and present some of their applications.

The remaining parts of the paper are organized as follows. In Sec.II, we give a brief review on the properties of heavy and light mesons' LCWFs. In Sec.III, we show some of their typical applications, i.e. the $B\to \pi $ TFFs, the determination of $|V_{\rm cb}|$ and the light meson-photon TFFs. The final section is reserved for a summary and outlook.

\section{Properties of the meson wavefunctions}

\subsection{B meson WFs}

In HQET, the B meson WFs $\tilde{\Psi}_{\pm}(t,z^2)$ can be defined in terms of the vacuum-to-meson matrix element of the nonlocal operators~\cite{grozin}:
\begin{widetext}
\begin{equation}\label{hqeteq}
\langle 0 | \bar{q}(z) \Gamma h_{v}(0) |\bar{B}(p) \rangle = -
\frac{i f_{B} M}{2} {\rm Tr} \Bigg[ \gamma_{5}\Gamma \frac{1 +
\slash\!\!\! v}{2} \!\!\!\times \Bigg\{ \tilde{\Psi}_{+}(t,z^2)-
\slash\!\!\! z \frac{\tilde{\Psi}_{+}(t,z^2)
 -\tilde{\Psi}_{-}(t,z^2)}{2t}\Bigg\} \Bigg],
\end{equation}
\end{widetext}
where $z^{\mu}=(0, z^{-}, \mathbf{z}_\perp)$, $z^{2}= - \mathbf{z}_{\perp}^{2}$, $v^{2} = 1$, $t=v\cdot z$ and $p^{\mu} = M v^{\mu}$ is the B meson 4-momentum with mass $M$, $h_{v}(x)$ denotes the effective $b$-quark field and $\Gamma$ is a generic Dirac matrix. The path-ordered gauge factors are implied between constituent quark fields. After doing the Fourier transformation, we have $\tilde{\Psi}_{\pm}(t,z^2) \to \Psi_{\pm}(\omega,z^2)$. The B meson DAs can be obtained by taking the LC limit to the B meson WFs, i.e. $\phi_\pm(\omega)\equiv \lim_{z^2\to 0}\Psi_{\pm}(\omega,z^2)$.

\subsubsection{B meson WFs in WW approximation}
\label{suba}

The equation of motion of the light spectator quark imposes a strong constraint on the B meson WFs. When ignoring the 3-particle Fock states' contributions, one obtains the WW-type B meson WFs. In this case, their DAs are~\cite{bwave,qiao0}:
\begin{eqnarray}
\phi^{WW}_{+}(\omega) &=& \frac{\omega}{2\bar\Lambda^2}\theta(\omega)
\theta(2\bar\Lambda-\omega), \\
\phi^{WW}_{-}(\omega) &=& \frac{2\bar\Lambda-\omega}{2\bar\Lambda^2}
\theta(\omega) \theta(2\bar\Lambda-\omega) ,
\end{eqnarray}
where $\bar{\Lambda} = \frac{iv\cdot \partial \langle 0| \bar{q} \Gamma h_{v} |\bar{B}(p) \rangle} {\langle 0| \bar{q} \Gamma h_{v} |\bar{B}(p) \rangle} $ is the B meson ``effective mass''. $\theta(\omega)$ is the unit step function, $\theta(\omega)=0$ for $\omega<0$ and $\theta(\omega)=1$ for $\omega\geq 0$. The transverse part of the B meson WFs is a zero-{\it th} normal Bessel function~\cite{bwave}. More explicitly, after applying Fourier transformation,
\begin{displaymath}
\tilde\Psi_{\pm}(\omega,\mathbf{k}_\perp) =\int \frac{d^2\mathbf{z}_{\perp}} {(2\pi)^2}\exp(-i \mathbf{k}_\perp\cdot\mathbf{z}_\perp)\Psi_{\pm}(\omega,z^2) ,
\end{displaymath}
the WW WFs can be written as
\begin{eqnarray}
\tilde\Psi^{WW}_{+}(\omega,\mathbf{k}_\perp) &=&
\frac{\phi^{WW}_+(\omega)}{\pi}\delta\left(\mathbf{k}_\perp^2-
\omega(2\bar\Lambda-\omega)\right) , \\
\tilde\Psi^{WW}_{-}(\omega,\mathbf{k}_\perp) &=&
\frac{\phi^{WW}_-(\omega)}{\pi}\delta\left(\mathbf{k}_\perp^2-
\omega(2\bar\Lambda-\omega)\right) .
\end{eqnarray}
This indicates that the transverse dependence of the B meson WW WF behaves as a $\delta$-function and the WFs' dependence on the transverse and longitudinal momenta is strongly correlated via a combined variable $\mathbf{k}_\perp^2/[\omega(2\bar\Lambda-\omega)]$. This off-shell-energy-like transverse dependence has also been observed by using the dispersion relation and the quark-hadron duality~\cite{halperin}, which in some sense is consistent with the BHL idea for the light meson WFs.

\subsubsection{B meson WFs in 3-particle Fock states}

To get the B meson WFs $\Psi_{\pm}(\omega,z^2)$ including the 3-particle Fock states, one needs to know the transverse properties of the 3-particle WFs~\cite{bwave2,qiao,qiao0}: $\Psi_{V}(\rho,\xi,z^2)$, $\Psi_{A}(\rho,\xi,z^2)$, $X_{A}(\rho,\xi,z^2)$ and $Y_A(\rho,\xi,z^2)$. An approximate solution for $\Psi_{\pm}(\omega,z^2)$ including the 3-particle Fock states can be obtained by taking the assumptions:

\noindent {\bf I}) The final $\Psi_{\pm}(\omega,z^2)$ have the same transverse momentum dependence, i.e.
\begin{equation}
\Psi_{\pm}[\omega, z^2]=\phi_{\pm}(\omega)\chi[\omega, z^2] ,
\end{equation}
and all 3-particle WFs have the same transverse momentum dependence $\chi^{(h)}(\rho,\xi,z^2)$, i.e.
\begin{eqnarray}
\Psi_{A}(\rho,\xi,z^2) &=& \Psi_{A}(\rho,\xi) \chi^{(h)}(\rho,\xi,z^2) , \\
\Psi_{V}(\rho,\xi,z^2) &=& \Psi_{V}(\rho,\xi) \chi^{(h)}(\rho,\xi,z^2) , \\
X_{A}(\rho,\xi,z^2) &=& X_{A}(\rho,\xi) \chi^{(h)}(\rho,\xi,z^2) , \\
Y_{A}(\rho,\xi,z^2) &=& Y_{A}(\rho,\xi)\chi^{(h)}(\rho,\xi,z^2),
\end{eqnarray}
with the boundary condition $\lim_{z^2\to 0}\chi^{(h)}(\rho,\xi,z^2)=1$.

\noindent {\bf II}) It is noted that main features of the 3-particle DAs are dominated by its first several moments, then one can assume that the relation among the first non-zero double moments of the 3-particle DAs are also satisfied by the 3-particle DAs themselves, e.g.
\begin{equation}
Y_A(\rho,\xi)\simeq\frac{X_A(\rho,\xi)- 3\Psi_A(\rho,\xi)}{4}\ .
\end{equation}

\noindent {\bf III}) The difference between $\Psi_V(\rho,\xi)$ and $\Psi_A(\rho,\xi)$ satisfies the relation~\cite{alex},
\begin{equation}
\Psi_V(\rho,\xi)-\Psi_A(\rho,\xi)=\frac{\lambda_{H}^2 -\lambda_{E}^2} {6\bar\Lambda^5} \rho\xi^2\exp
\left(-\frac{\rho+\xi}{\bar\Lambda}\right) ,
\end{equation}
where $\lambda_{E}$ and $\lambda_{H}$ are matrix elements of chromoelectric and chromomagnetic fields in B meson rest frame.

\begin{figure*}
\centering
\includegraphics[width=0.45\textwidth]{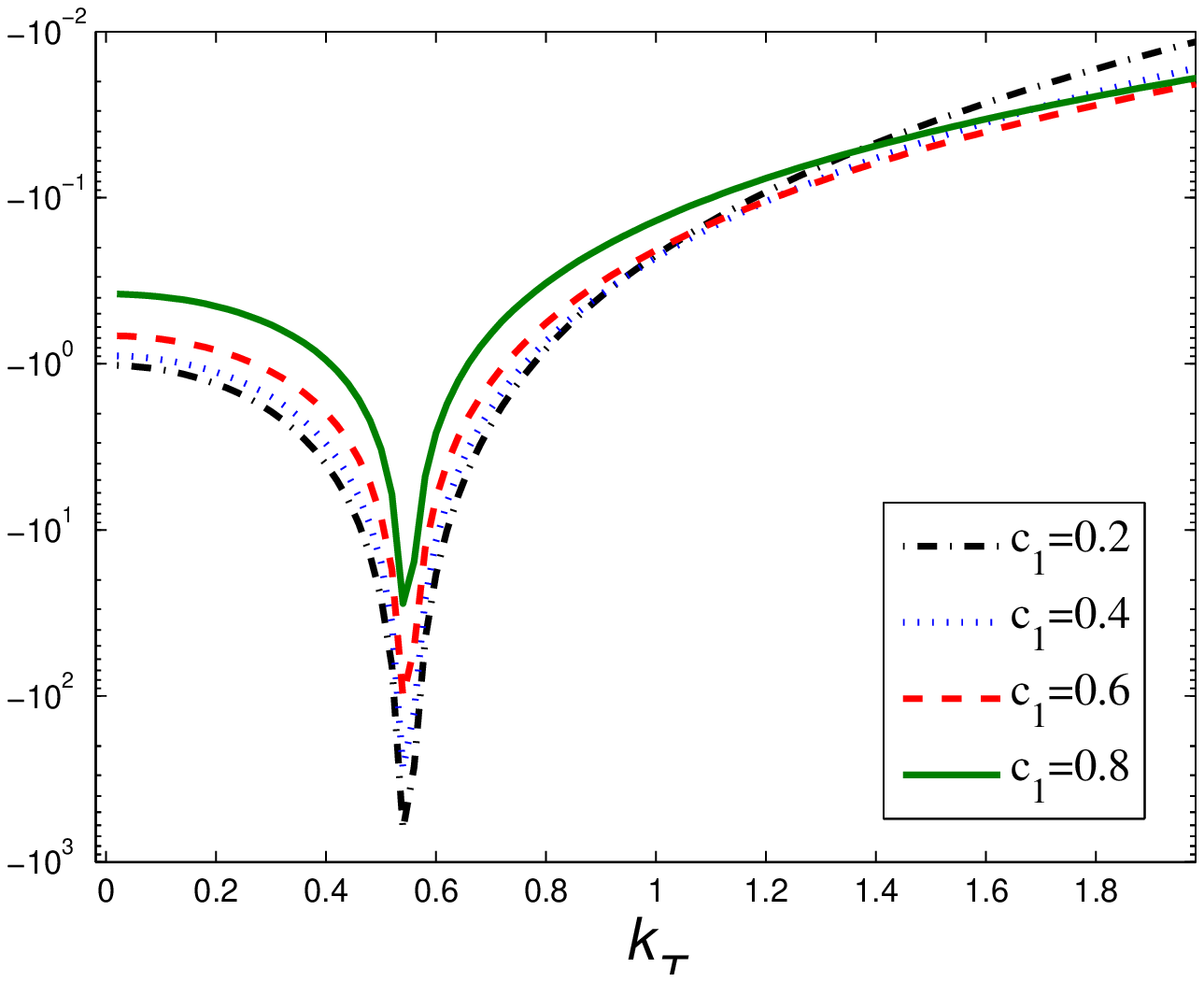}
\includegraphics[width=0.45\textwidth]{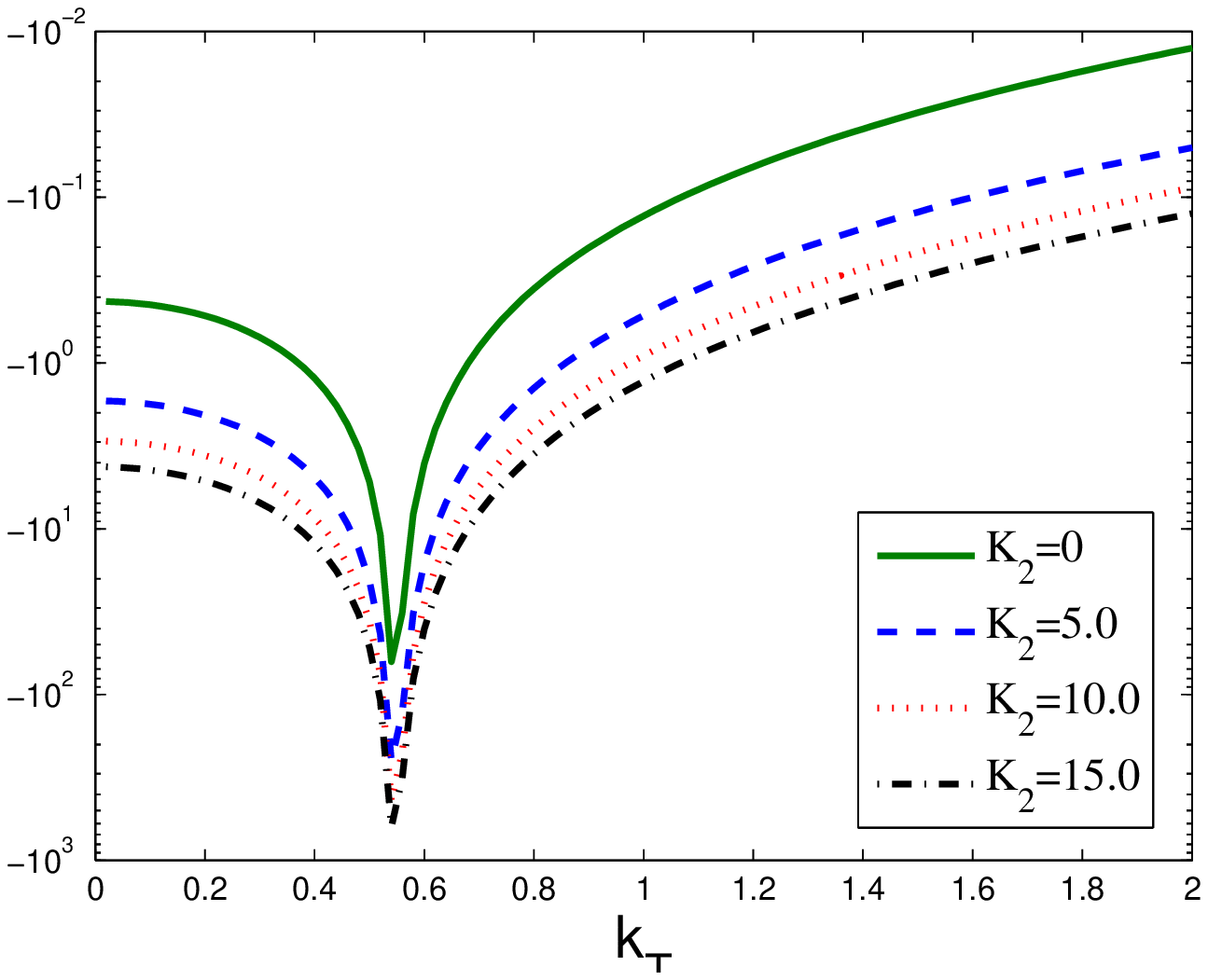}
\caption{The normalized transverse distribution $\tilde\chi(\omega,\mathbf{k}_\perp)$ ($k_T=|\mathbf{k}_\perp|$) with $\bar\Lambda=0.55$ GeV and $\omega=0.5$ GeV~\cite{bwave2}. The left diagram is for $K_2=1$ with $c_1=0.2$, $0.4$, $0.6$ and $0.8$, respectively; while the right diagram is for $c_1=0.6$ with $K_2=0$, $5.0$, $10.0$ and $15.0$, respectively. }
\label{fig2}
\end{figure*}

With those assumptions, one can obtain the normalized transverse distributions for the B meson WFs, which when transformed into the $\mathbf{k}_{\perp}$ space takes the following form~\cite{bwave2}:
\begin{eqnarray}
\tilde\chi(\omega,\mathbf{k}_\perp) &=& -\left(\frac{1} {\Gamma[1-c_1]\pi} +\frac{\Gamma[2-c_1]\sin[\pi c_1]}{\pi^2} K_2 \right) \nonumber\\
&&\frac{\Gamma[2-c_1]}{{((2\bar\Lambda - \omega ) \omega) } \left| 1 - \frac{k_T^2} {(2\bar\Lambda-\omega)\omega } \right|^{2-c_1}}, \label{momentum}
\end{eqnarray}
where $K_2$ is an input parameter, $c_1\in (0,1)$ and $k_T=|\mathbf{k}_\perp|$. It indicates that the B meson WF may be expanded to a hyperbola-like curve rather than a simple WW-like $\delta$-function. As an example, we put the transverse distributions of $\tilde\chi(\omega,\mathbf{k}_\perp)$ with fixed $\bar\Lambda=0.55$ GeV and $\omega=0.5$ GeV in Fig.(\ref{fig2}).

As for the B meson DAs $\phi_{\pm}(\omega)$, they have a quite complex form by considering the higher Fock state into consideration. By using the approximation, $\lambda_E^2 \simeq\lambda_H^2 \simeq2\bar\Lambda^2/3$, they have the following simpler forms,
\begin{equation}
\phi_+(\omega)=\frac{\omega}{\omega_0^2}\exp \left( -\frac{\omega}{\omega_0} \right), \;\;\; \phi_-(\omega)=\frac{1}{\omega_0}\exp \left(
-\frac{\omega}{\omega_0}\right),
\end{equation}
where $\omega_0=2\bar\Lambda/3$. The $\phi_{\pm}(\omega)$ agrees well with the model raised by Ref.\cite{grozin} within a QCD sum rule analysis.

\subsubsection{A simple model for B meson WFs}

To be more applicable, we propose a simple model for the B meson WFs with 3-particle Fock states' contributions. For convenience, we write the normalized B meson WFs in the compact parameter $b$-space (especially useful for the $k_T$-factorization approach~\cite{kt1,kt2,kt3,kt4}):
\begin{widetext}
\begin{eqnarray} \label{newmodel1}
\Psi_+(\omega,b) &=& \frac{\omega}{\omega_0^2}\exp \left(
-\frac{\omega}{\omega_0}\right) \Big(\Gamma[\delta] J_{\delta-1}
[\kappa] +(1-\delta)\Gamma[2-\delta] J_{1-\delta}[\kappa]\Big)\left(
\frac{\kappa}{2} \right)^{1-\delta} ,\\
\Psi_-(\omega,b) &=& \frac{1}{\omega_0}\exp \left(
-\frac{\omega}{\omega_0}\right)\Big(\Gamma[\delta] J_{\delta-1}
[\kappa] +(1-\delta)\Gamma[2-\delta] J_{1-\delta}[\kappa]\Big)\left(
\frac{\kappa}{2} \right)^{1-\delta} ,\label{newmodel2}
\end{eqnarray}
\end{widetext}
where $\omega_0=2\bar\Lambda/3$, $\kappa=\sqrt{\omega (2\bar\Lambda-\omega)}b $, $\delta\in(0,1)$ and $\lambda_E^2 \simeq\lambda_H^2 \simeq2\bar\Lambda^2/3$ is adopted. The range of $\omega$ is fixed within the range of $(0,2\bar\Lambda)$. When $\delta\to 1$, the transverse momentum dependence of the B meson WF returns to that of the WW-like B meson WF.

\subsection{Lighter meson WFs}

In this subsection, we sequentially present the leading-twist WFs for several typical lighter mesons, such as pion, $\eta_c$, kaon and D meson, respectively. We construct the WFs based on the BHL prescription. Because of the Wigner-Melosh rotation~\cite{wigner}, as a sound estimation, one also has to take the higher-order helicity components into consideration to construct the WF. As a subtle point, it is noted that such idea of constructing the light meson WFs can be conveniently extended to scalar or vector mesons, or to higher twist WFs. For example, the twist-3 WFs for pion and kaon following the same spirit can be found in Refs.\cite{pi4,kaon1}.

\begin{table}[htb]
\begin{center}
\begin{tabular}{ccccc}
\hline
~~~$\lambda_1\lambda_2$~~~ & ~~~$\uparrow\uparrow$~~~ &
~~~$\uparrow\downarrow$~~~ & ~~~$\downarrow\uparrow$~~~ & ~~~$\downarrow\downarrow$~~~ \\
\hline
$\chi^{\lambda_{1}\lambda_{2}}(x,\vec{k}_\perp)$ &
$-\frac{k_x- i k_y} {\sqrt{2}m_{qT}}$ & $\frac{m_q}{\sqrt{2}m_{qT}}$&
$-\frac{m_q}{\sqrt{2}m_{qT}}$ & $-\frac{k_x+i k_y}{\sqrt{2}m_{qT}}$ \\
\hline
\end{tabular}
\caption{Pion spin-space WF $\chi^{\lambda_{1}\lambda_{2}}(x,\vec{k}_\perp)$, where $\vec{k}_\perp=(k_x,k_y)$ and $m_q$ stands for the light constituent quark mass. The transverse mass $m_{qT}=\sqrt{m^{2}_q +k_\perp^2}$. }
\label{tab0}
\end{center}
\end{table}

Firstly, based on the BHL prescription, the pion leading-twist WF can be constructed as~\cite{TFF1,TFF2,TFF3,TFF4,TFF5,TFF6,TFF7}
\begin{eqnarray}
\Psi_\pi(x,\textbf{k}_\bot) = \sum_{\lambda_1 \lambda_2} \chi^{\lambda_1 \lambda_2}(x,\textbf{k}_\bot) \Psi^R_\pi(x,\textbf{k}_\bot), \label{WF_full}
\end{eqnarray}
where $\chi^{\lambda_1 \lambda_2}(x,\textbf{k}_\bot)$ stands for the spin-space WF, $\lambda_1$ and $\lambda_2$ being the helicity states of the two constitute quarks in pion. The $\chi^{\lambda_1 \lambda_2}(x,\textbf{k}_\bot)$ comes from Wigner-Melosh rotation~\cite{wigner}, whose explicit forms are presented in Table \ref{tab0}. $\Psi^R_\pi(x,\textbf{k}_\bot)$ indicates the spatial WF, whose $\textbf{k}_\bot$-dependent part can be constructed by using the connection between the rest frame WF $\Psi_{c.m.}(\textbf{q})$ and the LCWF $\Psi_{LC}(x,\textbf{k}_\bot)$, i.e.
\begin{eqnarray}
\Psi_{c.m.}(\textbf{q}^2) \longleftrightarrow \Psi_{LC} \left[ \frac{\textbf{k}_\bot^2 + m^2_q}{4x(1-x)} - m^2_q \right],
\end{eqnarray}
where $m_q\sim0.3$ GeV stands for the mass of the light constitute quarks of pion. As for the pure $x$-dependent part of $\Psi^R_\pi(x,\textbf{k}_\bot)$, we take $\varphi_\pi(x) = [1 + B_{\pi} \times C^{3/2}_2(2x-1)]$ with $B_{\pi} \sim a^{\pi}_2$, which dominates the WF longitudinal distribution. Then, the full form of the spatial WF can be obtained,
\begin{eqnarray}
\Psi^R_\pi(x,\textbf{k}_\bot) = A_{\pi} \varphi_\pi(x) \exp \left[ -\frac{\textbf{k}^2_\bot + m_q^2}{8\beta^2_\pi x(1-x)} \right], \label{WF_spatial}
\end{eqnarray}
where $A_{\pi}$ is the overall normalization constant. After integrating over the transverse momentum dependence, one obtains the pion DA,
\begin{widetext}
\begin{eqnarray}
\phi_\pi(x,\mu^2_0) = \frac{\sqrt{3} A_{\pi} m_q \beta_\pi}{2\pi^{3/2}f_\pi} \sqrt{x(1-x)} \varphi_\pi(x) \times \left\{ \textrm{Erf}\left[ \sqrt{\frac{m_q^2 + \mu_0^2}{8\beta^2_\pi x(1-x)}} \right] - \textrm{Erf}\left[ \sqrt{\frac{m_q^2}{8\beta^2_\pi x(1-x)}} \right] \right\}, \label{DA_model}
\end{eqnarray}
\end{widetext}
where $\mu_0$ is the factorization scale and $\textrm{Erf}(x)$ is the usual error function, $\textrm{Erf}(x) = \frac{2}{\sqrt{\pi}} \int^x_0 e^{-t^2} dt$.

Two constraints can be adopted to constrain the pion WF: (1) the normalization condition
\begin{eqnarray}
\int^1_0 dx \int \frac{d^2 \textbf{k}_\bot}{16\pi^3} \Psi_\pi(x,\textbf{k}_\bot) = \frac{f_\pi}{2\sqrt{6}};
\end{eqnarray}
(2) the sum rule derived from $\pi^0 \to \gamma\gamma$ decay amplitude
\begin{eqnarray}
\int^1_0 dx \Psi_\pi(x, \textbf{k}_\bot = \textbf{0}) = \frac{\sqrt{6}}{f_\pi},
\end{eqnarray}
where $f_{\pi}=130.41 \pm 0.03\pm0.20$ MeV~\cite{pdg}.

\begin{table}
\begin{center}
\begin{tabular}{cccccc}
\hline ~~~$B_{\pi}$~~~ & ~$A_{\pi} ({\rm GeV}^{-1})$~ & ~$\beta_{\pi} ({\rm GeV})$~ & ~$a^{\pi}_2(\mu_0^2)$~ & ~$a^{\pi}_4(\mu_0^2)$~  \\
\hline
~$0.00$~ & ~$24.80$~& ~$0.589$~ & ~$0.028$~ & ~$-0.027$~  \\
~$0.30$~ & ~$20.05$~& ~$0.672$~ & ~$0.364$~ & ~$-0.017$~  \\
~$0.60$~ & ~$16.46$~& ~$0.749$~ & ~$0.681$~ & ~$0.022$~   \\
\hline
\end{tabular}
\caption{Pion WF parameters for $m_q=0.30$ GeV~\cite{TFF7}. The second and fourth Gegenbauer moments are also presented as an explanation of pion DA behavior. It shows $B_\pi\sim a^{\pi}_{2}$, which indicates that $B_\pi$ dominantly determines the broadness of the longitudinal behavior of WF. }  \label{tab1}
\end{center}
\end{table}

\begin{figure}
\centering
\includegraphics[width=0.50\textwidth]{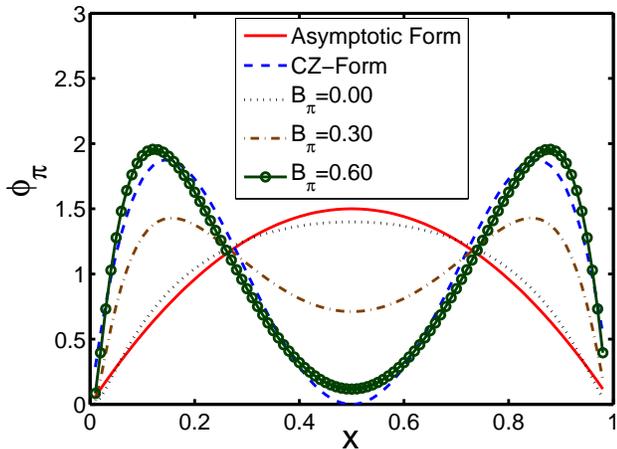}
\caption{The pion DA model $\phi_\pi(x,\mu_0^2)$ versus the parameter $B_\pi$~\cite{TFF4}. By varying $B_\pi$ from $0.00$ to $0.60$, $\phi_\pi(x,\mu_0^2)$ changes from asymptotic-like to CZ-like. } \label{phi}
\end{figure}

We put the pion WF parameters at the scale $\mu_0=1$ GeV in Table \ref{tab1}. When $B_\pi\in[0.00,0.60]$, such DA model can mimic the pion DA shapes from asymptotic-like~\cite{brodsky} to CZ-like~\cite{cz}. More explicitly, we put the DAs for $B_\pi=0.00$, $0.30$ and $0.60$ in Fig.(\ref{phi}), where as a comparison, the asymptotic DA and CZ-DA have also been present. If we have precise measurements for certain processes, then by comparing the theoretical estimations, one can fix the pion DA behavior.

Moreover, the pion DA can be run into any other scales by applying the evolution equation, i.e. to order ${\cal O}(\alpha_s)$, we have~\cite{brodsky}
\begin{widetext}
\begin{eqnarray}
&& x_1 x_2 \frac{\partial \tilde{\phi}_{\pi}(x_i,\mu^2)}{\partial \ln\mu^2} = C_F \frac{\alpha_s(\mu^2)}{4\pi} \times \left\{\int_0^1 [dy] V(x_i,y_i) \tilde{\phi}_{\pi}(y_i,\mu^2) - x_1 x_2 \tilde{\phi}_{\pi}(x_i,\mu^2) \right\}, \label{eq:evolution}
\end{eqnarray}
where $[dy]= dy_1 dy_2 \delta(1- y_1 -y_2)$, $\phi_{\pi}(x_i,\mu^2) = x_1 x_2 \tilde{\phi}_{\pi}(x_i,\mu^2)$ and
\begin{eqnarray}
V(x_i,y_i) &=& 2 x_1 y_2 \theta(y_1 - x_1) \left(\delta_{h_1\bar{h_2}}
+ \frac{\Delta}{(y_1 - x_1)} \right)  + (1 \leftrightarrow 2). \nonumber
\end{eqnarray}
\end{widetext}
The color factor $C_F=4/3$, $\delta_{h_1\bar{h_2}}=1$ for opposite $q$ and $\bar{q}$ helicities, and $\Delta\tilde{\phi}_{\pi}(y_i,\mu^2)=
\tilde{\phi}_{\pi}(y_i,\mu^2) - \tilde{\phi}_{\pi}(x_i,\mu^2)$. Practically, the above evolution (\ref{eq:evolution}) can be solved numerically or be solved by using the DA Gegenbauer expansion, which transforms the DA's scale dependence to the scale dependence of the Gegenbauer moments.

It is noted that, sometimes, a factor $1/x(1-x)$ or $\partial k_{z}/\partial x$ has also been introduced into the WF. It is noted that the factor $\partial k_{z}/\partial x$ comes from the transformation of the WF from the instant form to the light-cone form. The longitudinal component $k_{z}=(x-1/2)M_{0}$ with pion invariant mass $M_0 =\sqrt{{(\mathbf{k}^2_{\perp} +m^2_q)}/{x(1-x)}}$. These factors only slightly change the end-point behavior from the BHL-like terms and we only need to change the WF parameters properly.

\begin{figure}[tbh]
\includegraphics[width=0.50\textwidth]{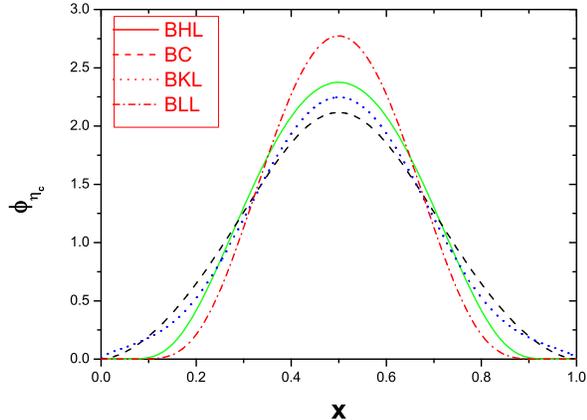}
\caption{A comparison of the $\eta_c$ DA $\phi_{\eta_c}(x,\mu_0^2)$~\cite{charm12} for $B_{\eta_c}=0$ with those of the BC model~\cite{charm2}, the BKL model~\cite{BKL2006} and the BLL model~\cite{BLL2007} at the initial scale $\mu_0=m_c$. } \label{fig1}
\end{figure}

Secondly, the $\eta_c$ WF can be obtained by applying the transition of $m_q\to m_c$ to Eq.(\ref{WF_full}). In addition to the normalization condition, we can use the condition $P_{\eta_c}\simeq 0.8$~\cite{wigner} to constrain the WF parameter, in which $P_{\eta_c}$ stands for the probability of finding the leading Fock state $|c\bar{c}>$ in the $\eta_c$ Fock state expansion, i.e.
\begin{equation}
P_{\eta_c} = \int_0^1 dx\int\frac{d^2 \vec{k_\perp}}{16\pi^3} |\Psi^{R}_{\eta_c}(x,\textbf{k}_\bot)|^2 .
\end{equation}
We set $\mu_0=m_c$ to be the initial scale for the non-perturbative distribution amplitude of the $\eta_c$-meson. For other choice of factorization scale, one can conveniently obtained its DA by using the above mentioned evolution equation (\ref{eq:evolution}). An an example, by taking the constituent quark mass $m_c=1.8$ GeV and the decay constant $f_{\eta_c}=0.335$ GeV~\cite{pdg}, we get $A_{\eta_c}=286~\mbox{GeV}^{-1}$ and $\beta_{\eta_c}=0.810~\mbox{GeV}^{-2}$ under the case of $B_{\eta_c}=0$. We present its DA in Fig.(\ref{fig1}), in which a comparison with those of Bondar-Chernyak (BC) model~\cite{charm2}, Bodwin-Kang-Lee (BKL) model~\cite{BKL2006} and Braguta-Likhoded-Luchinsky (BLL) model~\cite{BLL2007} has also been presented.

Thirdly, because the two constitute quarks are different, the leading-twist kaon WFs is slightly different from the case of pion. That is, the spatial part of the kaon WF $\Psi_{K}(x,\mathbf{k_\perp}) =\Psi^{R}_{K}(x,\mathbf{k_\perp}) \chi_{K}$ can be constructed as
\begin{widetext}
\begin{equation}
\Psi^{R}_{K}(x,\mathbf{k_\perp})=A_K [1+B_K C^{3/2}_1(2x-1)+C_K C^{3/2}_2(2x-1)] \times\exp \left[-b_K^2 \left(\frac{k_\perp^2+m_1^2}{x}+
\frac{k_\perp^2+m_2^2} {1-x} \right) \right] .
\end{equation}
\end{widetext}
Here the pure $x$-dependent part $\varphi_K(x)$ has also been expanded in Gegenbauer expansion, i.e.,
\begin{displaymath}
\varphi_K(x) = [1+B_K C^{3/2}_1(2x-1)+C_K C^{3/2}_2(2x-1)],
\end{displaymath}
where $m_1=m_u$ and $m_2=m_s$ for ${K^+}$, $m_1=m_d$ and $m_2=m_s$ for ${K^0}$. The length of the non-zero Gegenbauer terms depends on how well we know the Gegenbauer moments. Some other models have also been suggested for kaon WF, c.f. Refs.\cite{maboqiang,choi}, which are (numerically) consistent with the present model.

Because of different constitute quarks, kaon meson has non-zero first Gegenbauer moment. In the literature, the first and second Gegenbauer moments $a_1^K$ and $a^K_2$ have been studied by several approaches with high precision, such as Refs.~\cite{quark11,quark12,ballmoments,lcsr1,pballa1k, lenz,zwicky,lattice2,latt,instat,sumrule01,sumrule02}. So we keep two Gegenbauer terms in $\varphi_K(x)$. The spin-space part $\chi_{K}$ is more complex than the pionic case due to different constituent quark masses, whose explicit form can be found in Ref.\cite{kaon1}. Four parameters $A_K$, $B_K$, $C_K$ and $b_K$ can be determined by its DA's first two Gegenbauer moments $a^K_1$ and $a^K_2$, the constraint $\langle \mathbf{k}_\perp^2 \rangle^{1/2}_K \approx 0.350{\rm GeV}$~\cite{huangB1} and the normalization condition. As an example, by taking $a^K_1(1{\rm GeV})=0.05$~\cite{lcsr1} and $a^K_2(1GeV)=0.115$~\cite{sumrule01,sumrule02}, we have $A_{K}=12.55GeV^{-1}$, $B_{K}=0.0605$, $C_{K}=0.0348$ and $b_{K}=0.8706 GeV^{-1}$.

\begin{figure}[tb]
\begin{center}
\includegraphics[width=0.5\textwidth]{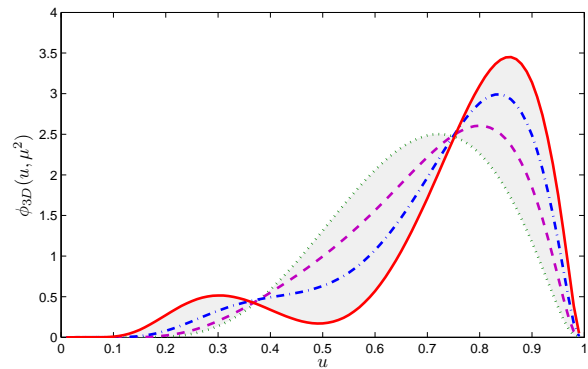}
\end{center}
\caption{The D meson DA $\phi_{D}(x,\mu^2_{0})$ at $\mu_0=1$ GeV with different $B_{D}$~\cite{charm10}. The dotted, the dashed, the dash-dot and the solid lines are for $B_{D}=0.00$, $0.20$, $0.40$ and $0.60$, respectively. } \label{DA_2}
\end{figure}

Fourthly, the $D$-meson WF $\Psi_D(x,\textbf{k}_\bot)$ can be constructed in the same way as the kaon WF, i.e. for the present case, $m_1=m_c$ and $m_2=m_d$. Because $m_c \gg m_d$, we shall have large non-zero first Gegenbauer moment $a_{1}^D$. As a simplified treatment, we find it is convenient to set the $x$-dependent part as $\varphi_D(x) = [1 + B_{D} \times C^{3/2}_{2} (2x-1)]$~\cite{charm10}, in which the value of $B_{D}$ is close to the second Gegenbauer moment, $B_{D} \sim a_{2}^D$, which basically determines the broadness of the longitudinal distribution. After integrating over the transverse momentum, we get the D meson DA. The parameter $B_{D}$ is a free parameter for $\phi_{D}$. As an example, we put the D meson DA $\phi_{D}$ with different $B_{D}$ in Fig.(\ref{DA_2}), in which we have set $B_{D}=0.00$, $0.20$, $0.40$ and $0.60$. By varying $B_{D}$ within a certain region, e.g. $[0.00,0.60]$, we can reproduce most of the D meson DAs suggested in the literature. For example, as $B_{D}=0.00$,it corresponds to the D meson wave function suggested in Ref.\cite{huangB1}. It is noted that by setting $B_{D} \in[0.00,0.40]$, we get the first Gegenbauer moment $a^{D}_{1}\sim[0.6,0.8]$, which is consistent with those suggested in the literature, and the second Gegenbauer moment $a^{D}_{2}\sim B_{D}$.

\section{Applications}

\subsection{$B\to\pi$ transition form factors}

\begin{figure}
\centering
\includegraphics[width=0.45\textwidth]{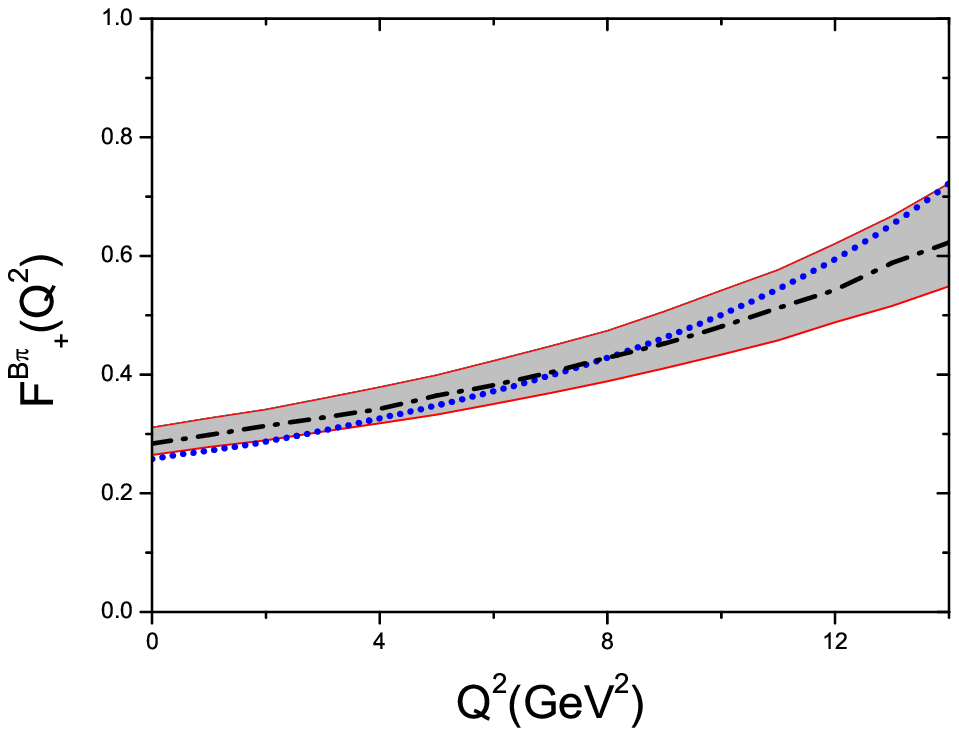}%
\hspace{0.2cm}
\includegraphics[width=0.45\textwidth]{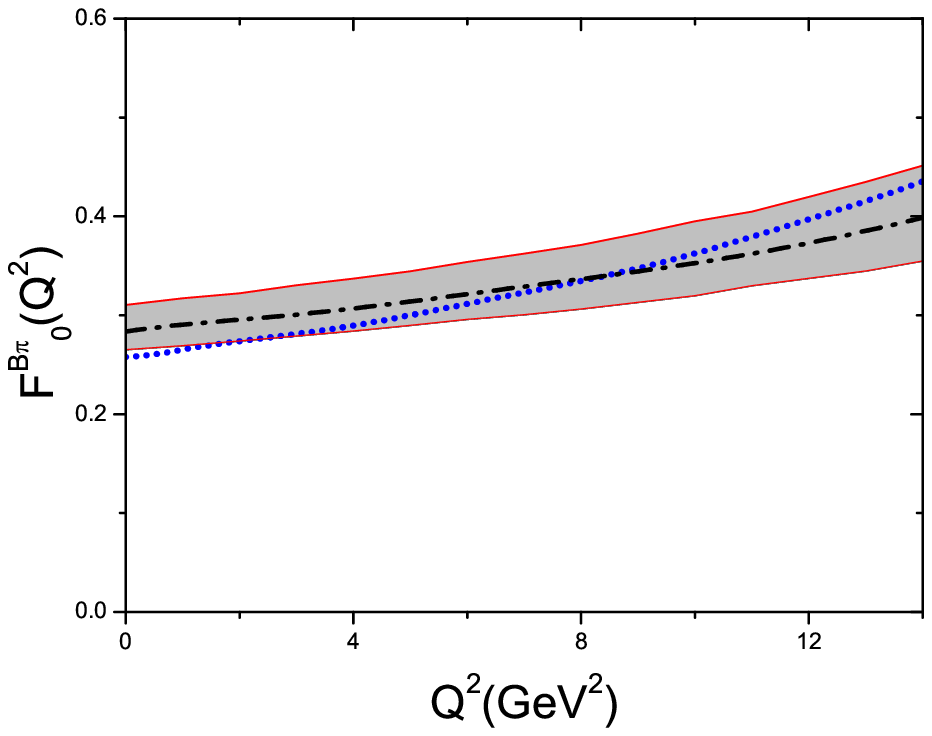}
\caption{The B meson TFFs $F^{B\pi}_+(Q^2)$ and $F^{B\pi}_0(Q^2)$~\cite{bwave2}. The upper (lower) edge of the shaded band is for $\delta=0.30$ ($\delta=0.25$), and the dash-dot line is for $\delta=0.27$. The dotted line is for the estimation of LCSRs~\cite{sumrule01,sumrule02}. $\bar\Lambda=0.55$ GeV. } \label{fbpi1}
\end{figure}

\begin{figure}
\centering
\includegraphics[width=0.45\textwidth]{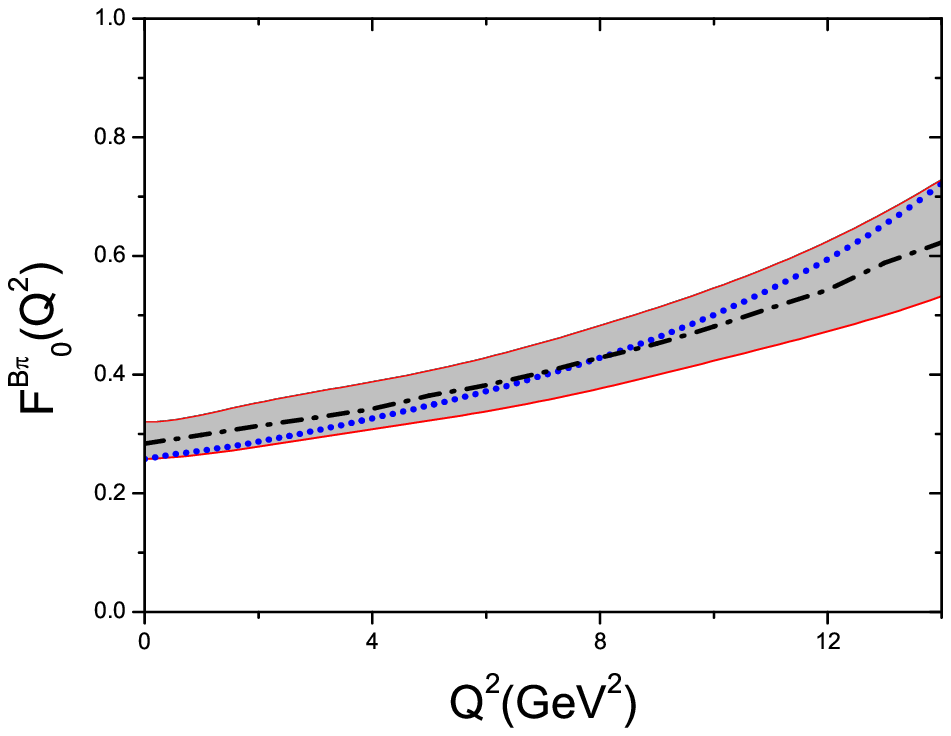}%
\hspace{0.2cm}
\includegraphics[width=0.45\textwidth]{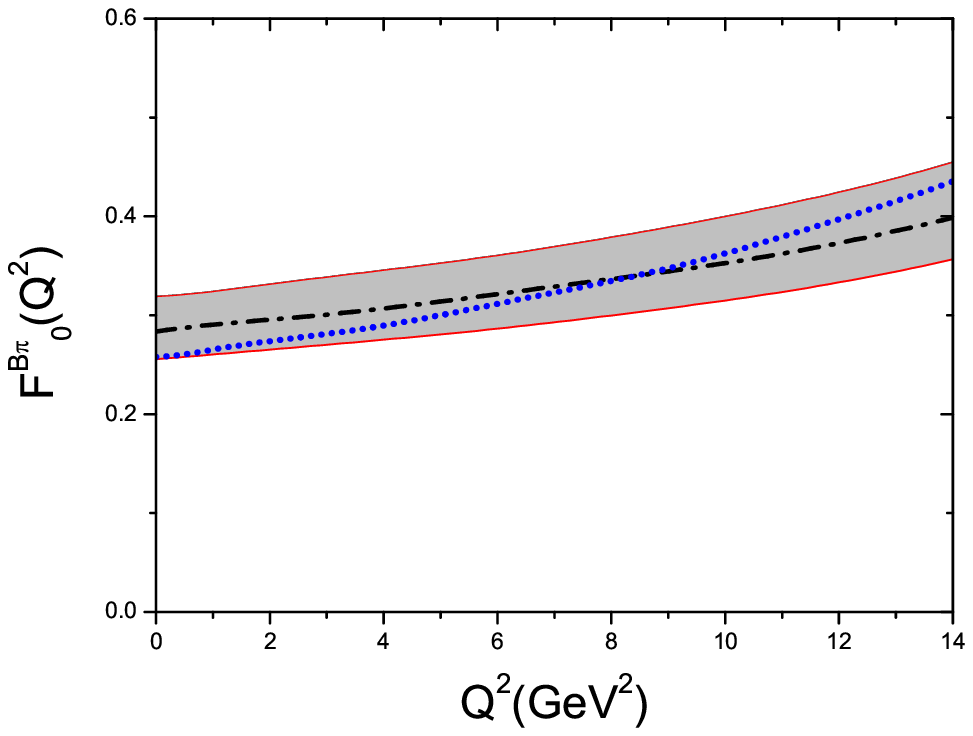}
\caption{The B meson TFFs $F^{B\pi}_+(Q^2)$ and $F^{B\pi}_0(Q^2)$~\cite{bwave2}. The upper (lower) edge of the shaded band is for $\bar\Lambda=0.52$ GeV ($\bar\Lambda=0.58$ GeV), the dash-dot line is for $\bar\Lambda=0.55$ GeV. The dotted line is for the estimation of LCSRs~\cite{sumrule01,sumrule02}. $\delta=0.27$. } \label{fbpi2}
\end{figure}

Applying the WW B meson WFs, we have shown that the $B\to\pi$ transition form factors (TFFs) calculated by using the $k_T$-factorization approach are consistent with those of QCD LCSRs and lattice QCD approaches~\cite{BPI_TFF_whole_region}, which are applicable in different $q^2$ regions. Results of Ref.\cite{BPI_TFF_whole_region} indicates that
\begin{itemize}
\item The negative contribution from $\bar{\Psi}_B=\frac{\Psi_{+} -\Psi_{-}}{2}$ is necessary to suppress the big contribution from $\Psi_B=\frac{\Psi_{+}+\Psi_{-}}{2}$ so as to obtain a reasonable total contributions. Thus, the properties of $\bar{\Psi}_B$ and $\Psi_B$ are important for an accurate estimation.
\item A better PQCD result (with its slope closest to the QCD LCSR results) can be obtained by carefully considering both the pionic twist-3 contributions and the contributions from the B meson WFs $\bar{\Psi}_B$ and $\Psi_B$. The pionic twist-3 WFs constructed via a similar way of leading twist WF can effectively suppress the end-point singularity and lead to a reliable twist-3 contributions.
\end{itemize}

As suggested in Eqs.(\ref{newmodel1},\ref{newmodel2}), by including 3-particle Fock states into B meson WFs, there are two unknown but universal parameters $\bar\Lambda$ and $\delta$ for the B meson WFs in compact parameter $b$-space. To see how the 3-particle Fock states affect the $B\to\pi$ TFFs, we calculate them within the $k_T$-factorization approach and show how $\bar\Lambda$ and $\delta$ affect the final estimations. Inversely, by comparing with the data, one can obtain reasonable regions for $\bar\Lambda$ and $\delta$~\cite{wurange}.

As a rough estimation, we show how the 3-particle Fock states affect the B meson decays. That is, a rough order estimation of $B\to\pi$ TFFs can be obtained by the first inverse moment of B meson DA $\phi_+(\omega)$, which satisfies
\begin{equation} \label{con4}
\Lambda_0=\int \frac{d\omega} {\omega}\phi_{+}(\omega)= \frac{1}{\lambda_B},
\end{equation}
where $\lambda_B=460 \pm 160$ MeV~\cite{alex,braun}. One may find that $\left(\Lambda^{(g)}_0 /\Lambda_0^{WW}\right)= (\bar\Lambda-\lambda_B)/\lambda_B \sim 0.2$~\cite{bwave2}, where $\Lambda^{(g)}_0$ is the first inverse moment of $\phi^{(g)}_+(\omega)$ and $\Lambda_0^{WW}$ is that of $\phi^{WW}_+(\omega)$. In Ref.\cite{charng}, the 3-particle contributions are estimated by attaching an extra gluon to the internal off-shell quark line, and then $(1/m_b) \sim 0.2$ power suppression is readily induced.

\begin{figure}
\centering
\includegraphics[width=0.45\textwidth]{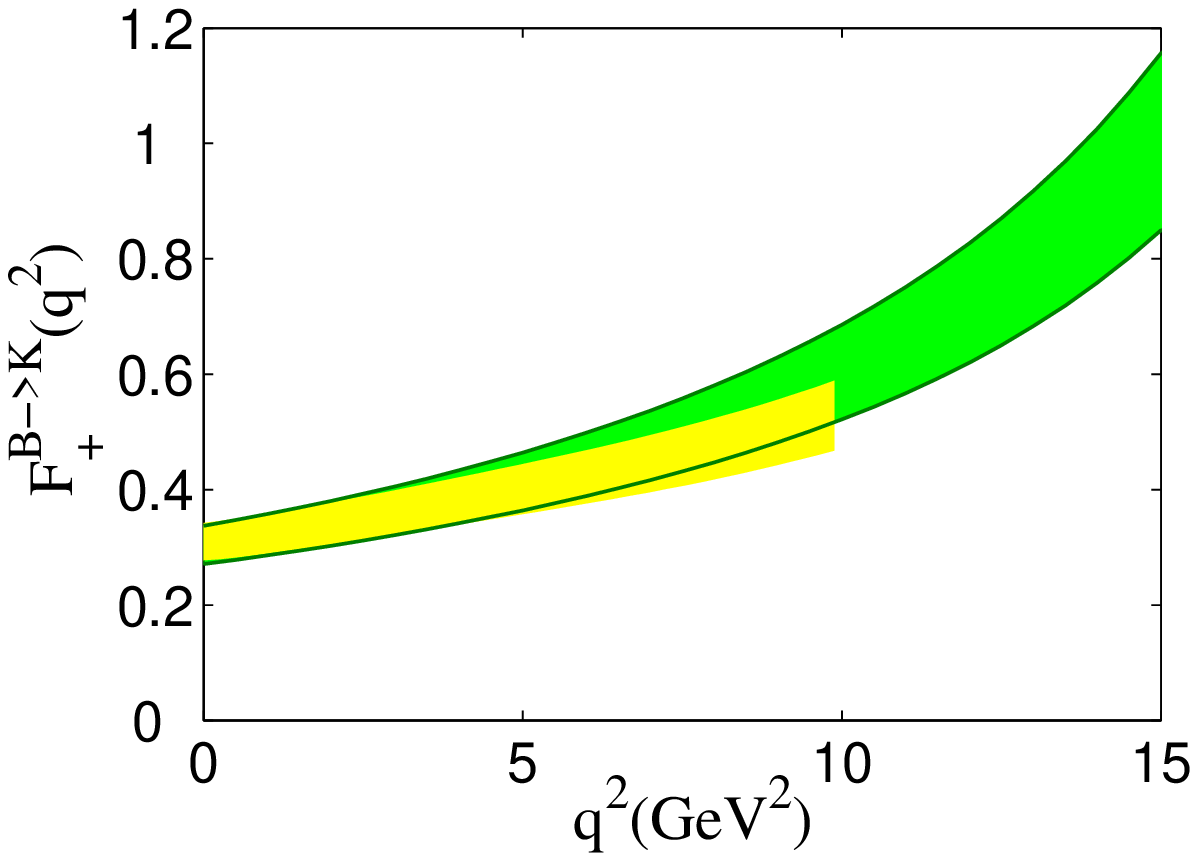}
\hspace{0.2cm}
\includegraphics[width=0.45\textwidth]{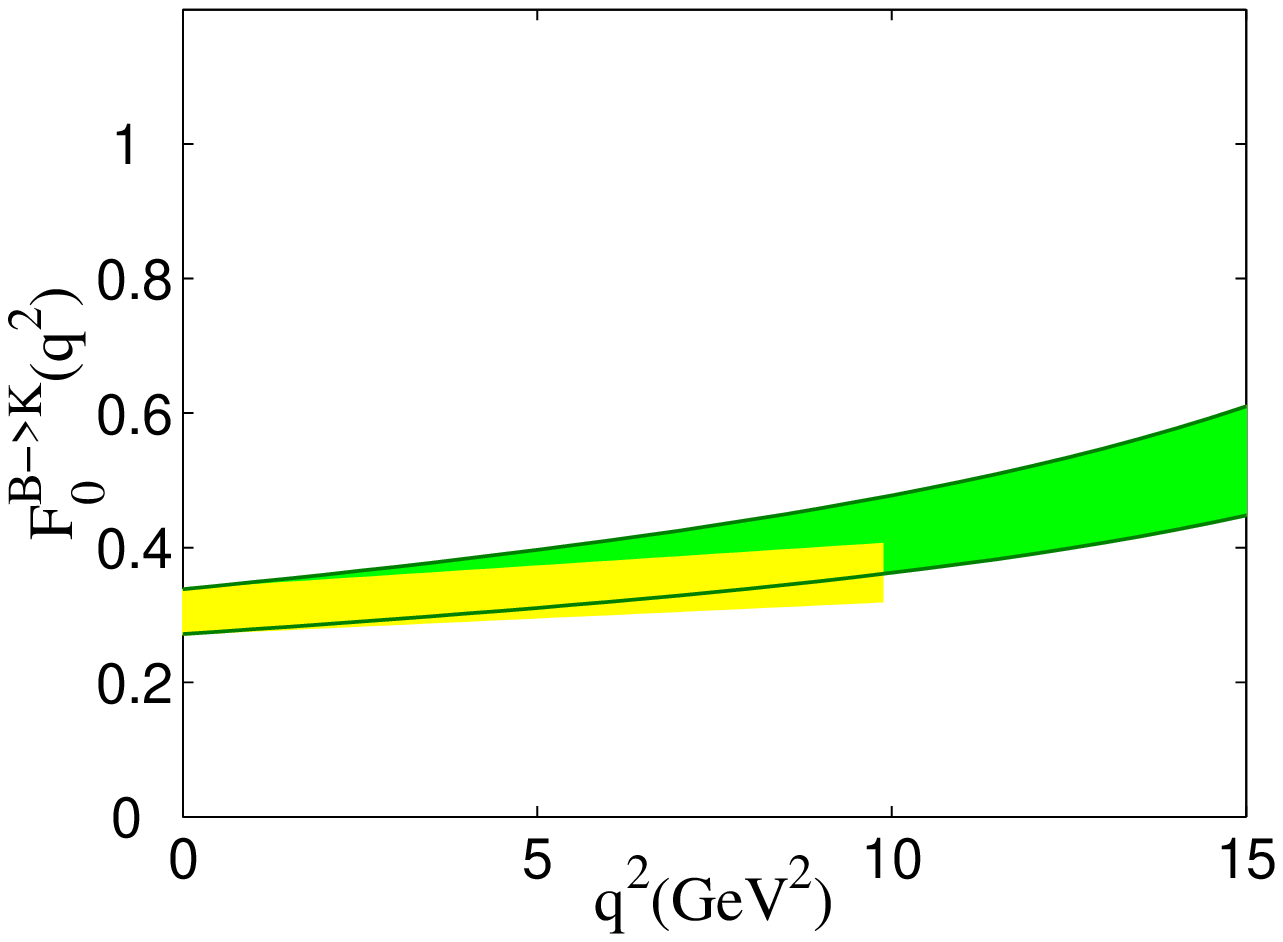}
\caption{A comparison of $F_{+,0}^{B\to K}(q^2)$ within the $k_T$ factorization and the LCSR~\cite{wurange}, where the lighter shaded band stands for $k_T$ factorization results for $\delta\sim (0.25,0.30)$ and $\bar\Lambda\sim (0.50GeV,0.55GeV)$, while, the thicker shaded band stands for LCSR results~\cite{sumrule01,sumrule02}. } \label{btok}
\end{figure}

By taking the B meson WFs as defined in Eqs.(\ref{newmodel1},\ref{newmodel2}) and a comparison of the estimations under the $k_T$-factorization and LCSR approaches, we make a discussion on the uncertainties of $B\to\pi$ TFF caused by $\delta$ and $\bar\Lambda$. To concentrate our attention on B meson WFs, we adopt the pion WF as that of Ref.\cite{BPI_TFF_whole_region}. We present the $B\to\pi$ TFFs $F^{B\pi}_{+,0}(Q^2)$ with fixed $\bar\Lambda=0.55$ GeV in Fig.(\ref{fbpi1}), while the cases with fixed $\delta=0.27$ are shown in Fig.(\ref{fbpi2}). Our results show that if the contribution from the 3-particle wavefunction is limited to be within $\pm 20\%$ of that of WW wavefunction with $Q^2\in (0,\sim 10GeV^2)$, then the possible range of $\delta$ and $\bar\Lambda$ are, $\delta\sim (0.25,0.30)$ and $\bar\Lambda\sim (0.50GeV,0.60GeV)$. This region has also been observed in Ref.\cite{wurange} by comparing with the pQCD and LCSRs estimations at $Q^2=0$, $F^{B\pi}_+(0) = 0.258\pm 0.031$ and $F^{B\pi}_T(0)=0.253\pm 0.028$~\cite{PB2005,bpisr}. This region can be further constrained by comparing $B\to K$ TFFs~\cite{wurange}, i.e. as shown by Fig.(\ref{btok}), we obtain $\delta\sim (0.25,0.30)$ and $\bar\Lambda\sim (0.50\;{\rm GeV}, 0.55\;{\rm GeV})$.

\subsection{Determination of $|V_{\rm cb}|$}

\begin{figure}[htb]
\begin{center}
\includegraphics[width=0.50\textwidth]{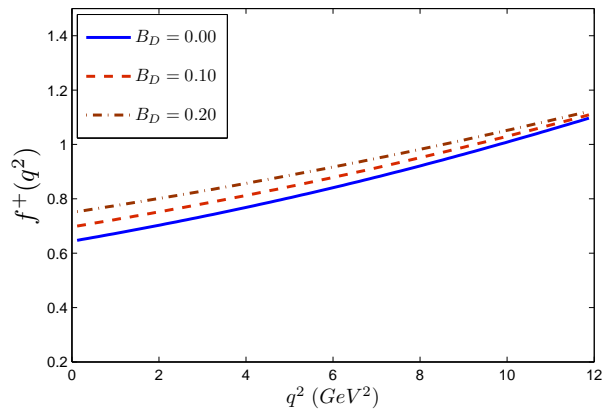}
\end{center}
\caption{The TFF $f^+(q^2)$ for D meson DA $\phi_{D}$ (\ref{phid}) with different choice of $B$~\cite{charm10}. The solid, the dashed and the dash-dot lines are for $B_D=0.00$, $0.10$ and $0.20$, respectively. } \label{fq2d}
\end{figure}

There are four $B\to D$ semi-leptonic processes that are frequently used to determine the CKM matrix element $\Vcb$, i.e. $B^0 \to D^- \ell^+ \nu_\ell$, $\bar B^0\to D^+ \ell^- \bar\nu_\ell$), $B^+ \to \bar D^0 \ell^+ \nu_\ell$ and $B^-\to D^0 \ell^- \bar\nu_\ell$. Schematically, we have
\begin{equation}
\frac{{\cal B}(B{\to}D\ell\bar{\nu}_\ell)}{\tau(B)} = \int_{0}^{(m_B  - m_D )^2 } {dq^2 \frac{{d\Gamma (B{\to}D\ell\bar{\nu}_\ell)}}{{dq^2 }}} .
\end{equation}
Here $\tau(B)$ stands for B meson lifetime and ${\cal B}(B{\to}D\ell \bar{\nu}_\ell)$ stands for the branching ratio of $B{\to}D\ell\bar{\nu}_\ell$, which are experimentally measurable. Under the limit $m_{\ell}\to 0$, the decay width takes the form~\cite{Heavy_Meson1,P.Ball_1}
\begin{equation}
\frac{d\Gamma}{dq^2}(B{\to}D\ell\bar{\nu}_\ell) =\frac{G_F^2\Vcb^2} {192 \pi^3m_B^3} \lambda^{\frac{3}{2}}(q^2)|f_{+}(q^2)|^2 ,
\end{equation}
where $\lambda(q^{2}) = (m_B^2  + m_D^2  - q^2)^2 - 4 m_B^2 m_D^2$ and $f_{+}(q^2)$ has been investigated by using several approaches, such as the lattice QCD~\cite{Lattice_1}, the pQCD~\cite{charm1} and the QCD LCSR~\cite{charm10,BDsr1,charm4,BDsr3,BDsr4}. Especially, a recent improved LCSR determination up to twist-4 accuracy has been presented in Ref.\cite{charm10}. By using the improved LCSR with a chiral current correlator, the most uncertain twist-3 DAs make no contributions and twist-4 parts also provides small contributions about several percent, so the reliability of LCSR estimation can be enhanced to a large degree. This inversely makes the $B\to D$ semi-leptonic decays be good places for testing different models for the leading twist D meson LCDA. Based on the D meson WF modeled in the last section, we get its DA as,
\begin{eqnarray}
\phi_{D}(x,\mu_0^2 )&=& \frac{A_{D} \sqrt{6x\bar{x}} Y}
{{8\pi ^{3/2} f_D b_{D} }}[1+B_D\; C_2^{3/2}(x-\bar x)]\times \nonumber\\
&&\exp \bigg[  - b_{D} ^2\frac{{x m_d^2  + \bar{x} m_c^2  - {\rm{Y}}^2 }}{{x\bar{x}}}\bigg] \times \nonumber\\
&& \left[ {{\rm{Erf}}\Big( {\frac{{b_{D}\sqrt{ {\mu_0 ^2  + {\rm{Y}}^2 } }}}{{\sqrt {x\bar{x}} }}} \Big) - {\rm{Erf}}\Big( {\frac{{b_{D} {\rm{Y}}}}{{\sqrt {x\bar{x}} }}} \Big)} \right], \label{phid}
\end{eqnarray}
where $A_{D}$, $B_D$ and $b_{D}$ are undetermined parameters. The error function ${\rm{Erf}}(x)$ is defined as $ {\rm{Erf}}(x)=2\int^x_0{\exp({-t^2})dt} /\sqrt{\pi} $, ${\rm{Y}} = x m_d+\bar{x}m_c$ and $\bar{x} = 1 - x $. Some of its typical behaviors are presented in Fig.(\ref{DA_2}). The TFFs for $\phi_{D}$ with $B_D=0.00$, $0.10$ and $0.20$ are presented in Fig.(\ref{fq2d}). It shows that $f^+(q^2)$ increases with the increment of $B_D$, which is consistent with the trends shown in Fig.(\ref{DA_2}) that a bigger $B_D$ leads to a weaker suppression in the end-point region $(x\to 0)$, and shall result in a larger estimation on $f^{+}(q^2)$.

\begin{table}[htb]
\begin{center}
\begin{tabular}{c  c  c  }
\hline
~~~$B_D$~~~ & ~~~$B^0/\bar{B}^0$-type~~~ & ~~~$B^{\pm}$-type~~~ \\
\hline
0.00 & $41.28 {^{+5.68}_{-4.82}}~ {^{+1.13}_{-1.16}}$  & $40.44 {^{+5.56}_{-4.72}}~ {^{+1.06}_{-1.09}}$ \\
0.10 & $39.01 {^{+5.25}_{-4.59}}~ {^{+1.06}_{-1.09}}$  & $38.22 {^{+5.14}_{-4.50}}~ {^{+1.01}_{-1.03}}$ \\
0.20 & $36.96{^{+ 4.98}_{- 4.43}}~ {^ {+ 1.01}_{- 1.04}}$  & $36.21 {^{+4.88}_{-4.34}}~ {^{+0.95}_{-0.98}}$ \\
\hline
\end{tabular}
\caption{The value of $\Vcb$ in unit $10^{-3}$~\cite{charm10}. The central values for $\Vcb$ are obtained by setting all inputs to be their central values. The errors are calculated by theoretical and experimental errors on all inputs. }  \label{Vcb_results}
\end{center}
\end{table}

We discuss the variation of $\Vcb$ by taking $\phi_{D}$ with several choices of $B_D$, in which the first (second) uncertainty comes from the squared average of the mentioned theoretical (experimental) uncertainties. The experimental uncertainty comes from the lifetime and the decay ratio of the mentioned processes~\cite{pdg}. The results are put in Table \ref{Vcb_results}. It is noted that the value of $\Vcb$ decreases with the increment of $B_D$. The matrix element $\Vcb$ and its uncertainties have been studied by using two types of processes, e.g. the $B^0/\bar{B}^0$-type ($B^0 \to D^- \ell^+ \nu_\ell$ and $\bar B^0\to D^+ \ell^- \bar\nu_\ell$) and the $B^{\pm}$-type ($B^+ \to \bar D^0 \ell^+ \nu_\ell$ and $B^-\to D^0 \ell^- \bar\nu_\ell$). For the case of $B_D$=0, by adding the errors from the experimental and theoretical uncertainty sources, we obtain $|V_{cb}|(B^0/\bar{B}^0-{\rm type})=(41.28 {^{+6.81}_{-5.98}}) \times 10^{-3}$ and $|V_{cb}|(B^{\pm}-{\rm type})=(40.44 {^{+6.54}_{-5.72}}) \times 10^{-3}$. As a weighted average of these two types we obtain,
\begin{equation}
|V_{cb}| = (40.84\pm3.11)\times 10^{-3} , \;\;(B_D =0.00)
\end{equation}
where the error stands for the standard derivation of the weighted average. Similarly, we have
\begin{eqnarray}
|V_{cb}| &=& (39.08\pm3.03)\times 10^{-3} , \;\;(B_D =0.10) , \\
|V_{cb}| &=& (37.59\pm2.89)\times 10^{-3} , \;\;(B_D =0.20) .
\end{eqnarray}

\begin{figure}[htb]
\begin{center}
\includegraphics[width=0.5 \textwidth]{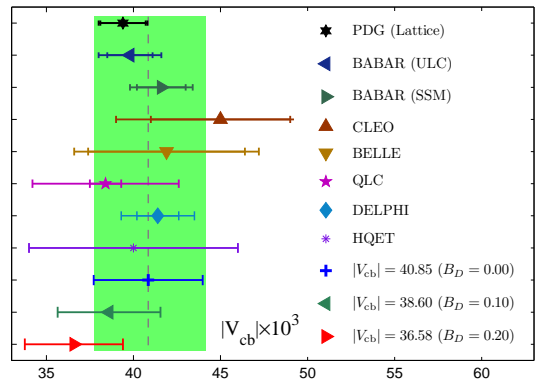}
\end{center}
\caption{A comparison of $\Vcb$ with experimental and theoretical predictions~\cite{charm10}. Our estimations for $B_D=0.00$, $0.10$ and $0.20$ are presented, where as a comparison, the results of BABAR~\cite{expV1}, PDG (Lattice)~\cite{pdg}, CLEO~\cite{expV3}, Belle~\cite{expV2}, QLC~\cite{Lattice_1}, DELPHI~\cite{DELPHI} and HQET~\cite{HQET} are also presented. }\label{Vcb_22}
\end{figure}

A comparison of $\Vcb$ with experimental and theoretical predictions is put in Fig.(\ref{Vcb_22}), in which our estimations for $B_D=0.00$, $0.10$ and $0.20$ are presented. Through a comparison with the experimental data, our present estimation for $\Vcb$ shows a good agreement with the BABAR, CLEO and Belle estimates. It shows that to compare with the data, we need a small value for $B_D$.

\subsection{Light meson-photon transition form factors}

Firstly, we present the results for the pion-photon TFF. The pion-photon TFF provides the simplest example for the perturbative analysis to exclusive process~\cite{TFFas}. To explain the abnormal large $Q^2$ behavior observed by the BABAR Collaboration in 2009~\cite{TFF_BABAR}, many works have been done, e.g. by the pQCD~\cite{TFF4,TFF5,TFF2010_Q0,TFF2011_MPA} or by LCSR~\cite{TFF2009_LCSR,TFF2011_LCSR1,TFF2011_LCSR2}. However, last year, the Belle Collaboration released their new analysis~\cite{TFF_BELLE}, which  dramatically different from those reported by BABAR Collaboration, but likely to agree with the asymptotic behavior estimated by Ref.\cite{TFFas}. Many attempts have been tried to clarify the situation~\cite{DA_fbpi,TFF6,TFF2012_LCSR1,TFF2012_LCSR2,TFF2012_Q0,TFF2013_LCSR}.

Generally, the pion-photon TFF $F_{\pi\gamma}(Q^2)$ can be written as a sum of the valence quart part $F_{\pi\gamma}^{(V)}(Q^2)$ and the non-valence quark part $F_{\pi\gamma}^{NV} (Q^2)$~\cite{TFF2007,TFF4,TFF5,TFF6}:
\begin{eqnarray}
F_{\pi\gamma}(Q^2) = F_{\pi\gamma}^{(V)}(Q^2) + F_{\pi\gamma}^{(NV)}(Q^2) .
\label{TFF_total}
\end{eqnarray}
The valence quark part $F^{(V)}_{\pi\gamma}(Q^2)$ indicates the pQCD calculable leading Fock-state contribution, which dominates the TFF when $Q^2$ is large. The non-valence quark part $F^{(NV)}_{\pi\gamma}(Q^2)$ is related to the non-perturbative higher Fork-states contributions, which can be estimated via a proper phenomenological model. Both the analytic expressions for $F_{\pi\gamma}^{(V)}(Q^2)$ and $F_{\pi\gamma}^{NV} (Q^2)$ can be found in Ref.\cite{TFF4}.

\begin{figure}[tb]
\centering
\includegraphics[width=0.50\textwidth]{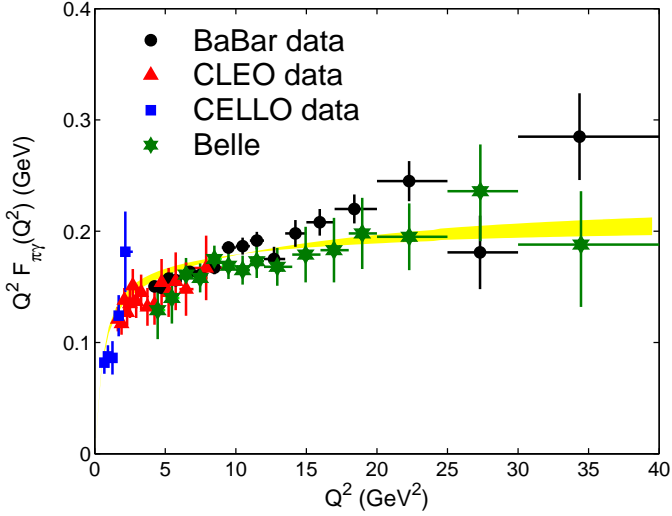}
\caption{$Q^2 F_{\pi\gamma}(Q^2)$ with our WF model (\ref{WF_full},\ref{WF_spatial}) by varying the model parameter $B_\pi$ within the region $[0.01,0.14]$~\cite{TFF7}. The shaded band is our theoretical estimation. The experimental data are taken from Refs.\cite{TFF_CELLO,TFF_CLEO01,TFF_CLEO02,TFF_BABAR,TFF_BELLE}.} \label{fig_PPTFF}
\end{figure}

Taking all input parameters to be the same as those adopted in Ref.\cite{TFF7}, we draw the pion-photon TFF $F_{\pi\gamma}(Q^2)$ in Fig.(\ref{fig_PPTFF}). The upper and lower borderlines correspond to $B_\pi=0.01$ and $0.14$ respectively. It shows that in the small $Q^2$ region, $Q^2\lesssim 15~GeV^2$, the pion-photon TFF can explain the CELLO~\cite{TFF_CELLO}, CLEO~\cite{TFF_CLEO01,TFF_CLEO02}, BABAR~\cite{TFF_BABAR} and Belle~\cite{TFF_BELLE} experimental data simultaneously. While for the large $Q^2$ region, our present estimation favors the Belle data and disfavors the BABAR data. If taking $B_\pi\in[0.01,0.42]$, the calculated curve for the pion-photon TFF with the upper limit of the parameter ($B_\pi=0.42$) will be between the Belle and BABAR data. It is expected that BABAR and Belle can obtain more accurate and consistent data in the future, then the behavior of the pion DA can be further determined completely.

\begin{figure}
\centering
\includegraphics[width=0.45\textwidth]{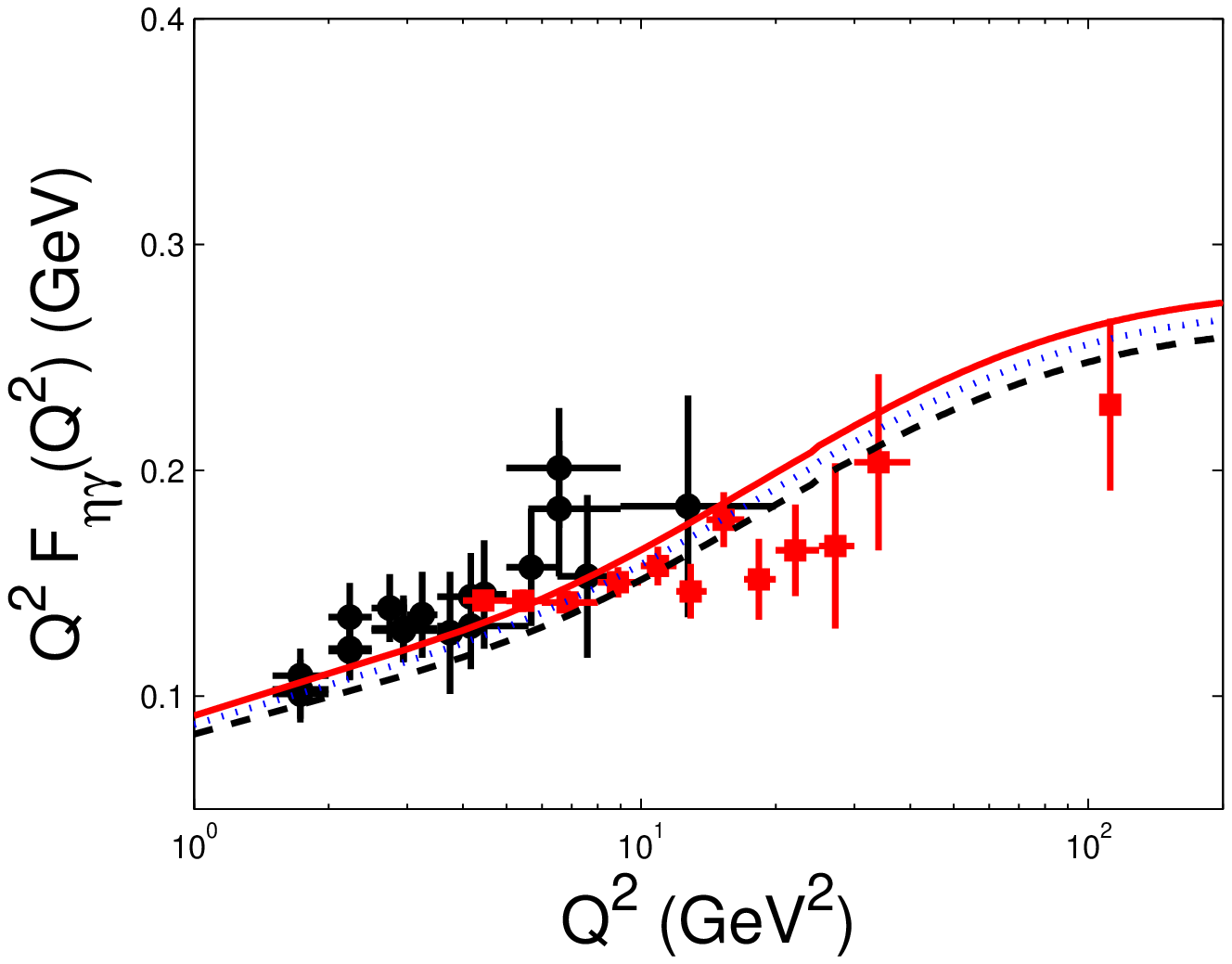}
\includegraphics[width=0.45\textwidth]{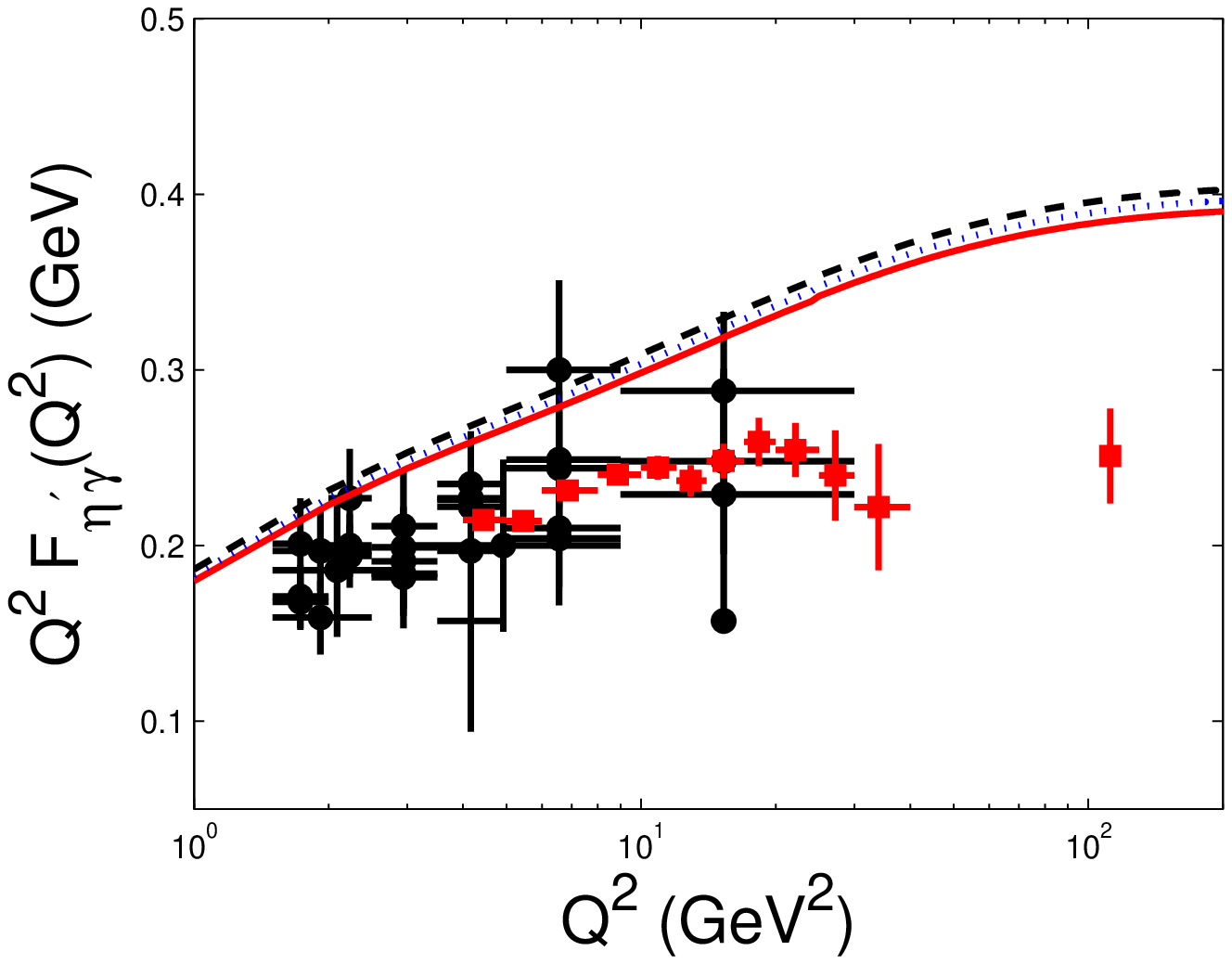}
\caption{$\eta$-$\gamma$ and $\eta'$-$\gamma$ TFFs $Q^{2}F_{\eta\gamma}(Q^2)$ and $Q^{2}F_{\eta'\gamma}(Q^2)$ with fixed $B_{\eta,\eta'}=0.3$ and $m_q=0.30$ GeV and $m_s=0.45$ GeV~\cite{TFF5}. The solid, the dotted and the dashed lines are for $\phi=39.0^{\circ}$, $39.5^{\circ}$ and $40.0^{\circ}$, respectively. The experimental data are taken from Refs.\cite{TFF_CLEO01,TFF_CLEO02,babarold,babareta}. } \label{phiuncern}
\end{figure}

Secondly, we discuss the $\eta$-$\gamma$ and $\eta'$-$\gamma$ TFFs. Under the quark-flavor mixing scheme, the $\eta$-$\gamma$ and $\eta'$-$\gamma$ TFFs are related with $F_{\eta_{q}\gamma}(Q^2)$ and $F_{\eta_{s}\gamma}(Q^2)$ through the following equations~\cite{TFF5}
\begin{eqnarray}
F_{\eta\gamma}(Q^2)&=&F_{\eta_q\gamma}(Q^2)\cos\phi -F_{\eta_s\gamma}(Q^2)\sin\phi \label{ffeta} \\
F_{\eta'\gamma}(Q^2)&=&F_{\eta_q\gamma}(Q^2)\sin\phi +F_{\eta_s\gamma}(Q^2)\cos\phi , \label{ffetap}
\end{eqnarray}
where $\phi$ is the mixing angle, which as a weighted average of experimental results~\cite{feldmann,bes,kroll}, is about $39.5^{\circ}\pm0.5^{\circ}$. Thus these two TFFs can also be adopted for determining the $\eta_q$ and $\eta_s$ WFs, which can be constructed in a similar way of pion WF. Due to $\eta$-$\eta'$ mixing, we need to consider these two TFFs simultaneously. The curves of $Q^{2}F_{\eta\gamma}(Q^2)$ and $Q^{2}F_{\eta'\gamma}(Q^2)$ for $\phi=39.5^{\circ}\pm 0.5^{\circ}$ are presented in Fig.(\ref{phiuncern}). With the increment of $m_q$ and $m_s$, the values of $Q^{2}F_{\eta\gamma}(Q^2)$ and $Q^{2}F_{\eta'\gamma}(Q^2)$ increase in lower $Q^2$ region and decrease in higher $Q^2$ region. And it is found that $Q^{2} F_{\eta\gamma}(Q^2)$ decreases with the increment of $\phi$, while $Q^{2}F_{\eta'\gamma}(Q^2)$ increases with the increment of $\phi$. By shifting $\phi$ to a smaller value $\sim 38^{\circ}$, one can explain them within $Q^2<20$ GeV$^2$ consistently~\cite{whtheta}, which can not explain the newly BABAR data on $\eta$-$\gamma$ and $\eta'$-$\gamma$ for larger $Q^2>20$ GeV$^2$. A proper intrinsic charm may have some help to explain the abnormally large production of $\eta'$. It has been shown that the experimental data disfavors a larger portion of charm component as $|f_{\eta'}^c| \gtrsim 50$ MeV~\cite{TFF5,whtheta}.

\begin{figure}
\centering
\includegraphics[width=0.45\textwidth]{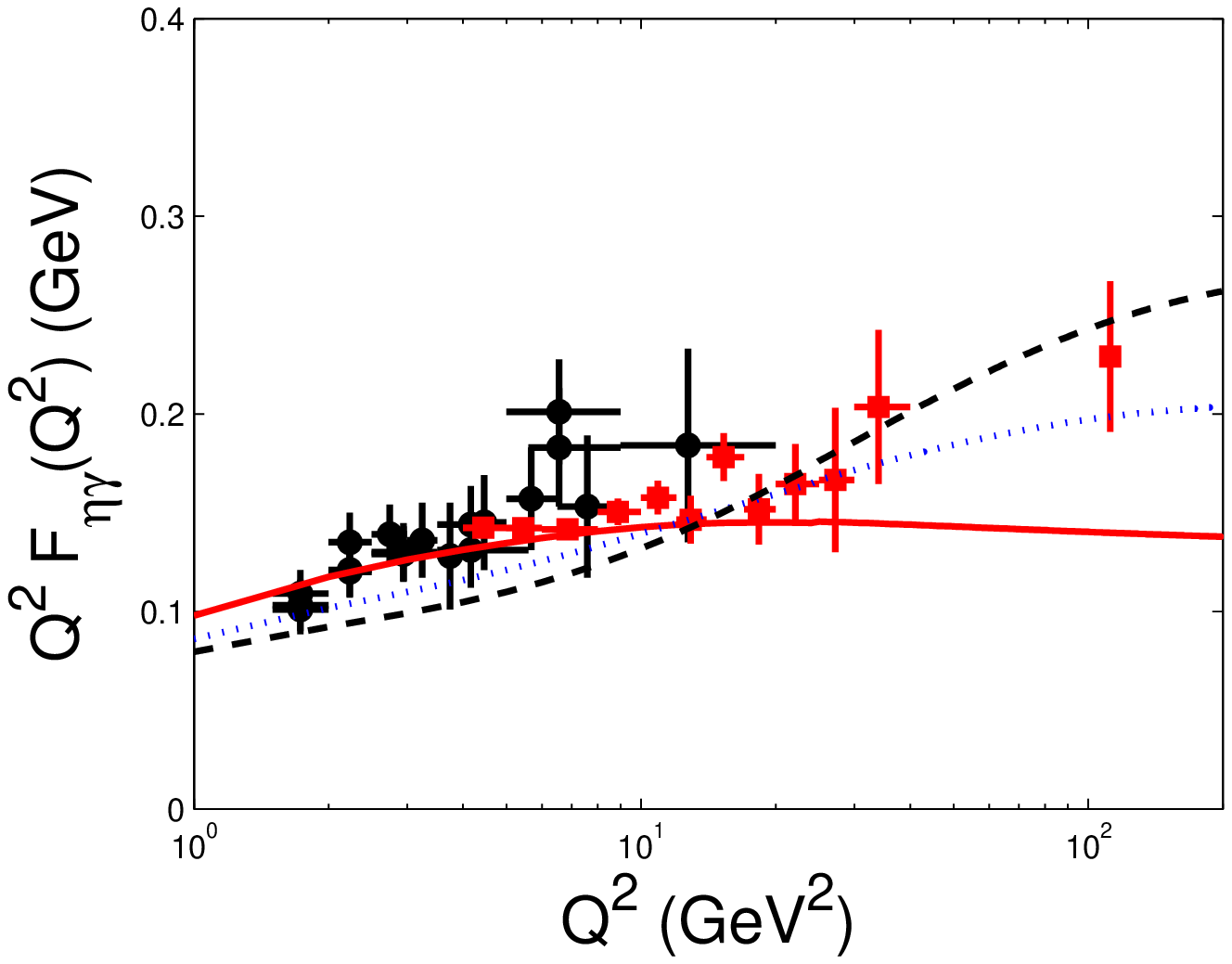}
\hspace{0.1cm}
\includegraphics[width=0.45\textwidth]{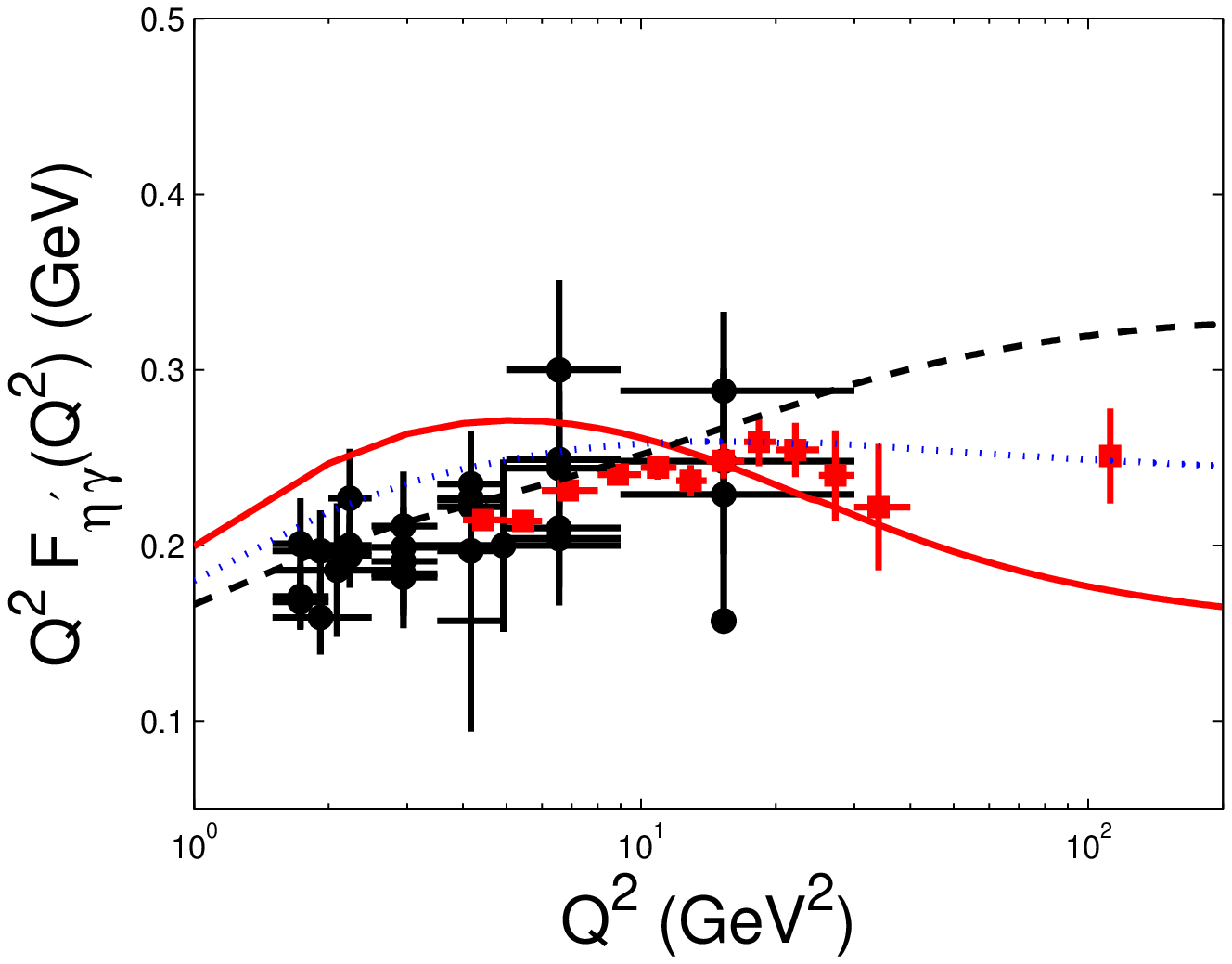}
\caption{$\eta$-$\gamma$ and $\eta'$-$\gamma$ TFFs $Q^{2}F_{\eta\gamma}(Q^2)$ and $Q^{2}F_{\eta'\gamma}(Q^2)$~\cite{TFF5}, where $f^{c}_{\eta'}=-30$ MeV, $m_q=0.30$ GeV, $m_s=0.45$ GeV and $m_c=1.50$ GeV. The solid, the dotted and the dashed lines are for $B_{\eta,\eta'}=0.0$, $0.30$ and $0.60$, respectively. The experimental data are taken from Refs.\cite{TFF_CLEO01,TFF_CLEO02,babarold,babareta}. } \label{charmB}
\end{figure}

More over, we present the results for $\eta$-$\gamma$ and $\eta'$-$\gamma$ TFFs with fixed $f^{c}_{\eta'}=-30$ MeV in Fig.(\ref{charmB}), where $B_{\eta,\eta'}=0.0$, $0.30$ and $0.60$ respectively. It shows that with proper charm component $f^{c}_{\eta'}\sim -30$ MeV and $B_{\eta,\eta'}\sim 0.30$, the experimental data on $Q^{2}F_{\eta\gamma}(Q^2)$ and $Q^{2}F_{\eta'\gamma}(Q^2)$ can be consistently explained. A moderate DA with $B_{\eta,\eta'}\sim 0.30$, corresponding to the second Gegenbauer DA moment around $0.35$ for $\eta_q$ and $\eta_s$ DAs.

\section{Summary and outlook}

The heavy and light meson WFs are nonperturbative but universal physical quantities in high energy processes and are important components for applying pQCD factorization theory. A better understanding of LCWF may also be helpful for solving the factorization scale setting problem following the idea of the principle of maximum conformality to solve renormalization scale ambiguity~\cite{pmc1,pmc2,pmc3,pmc4}. In the present paper, we have reviewed recent progresses on constructing the B meson and light pseudoscalar WFs and have presented some of their applications. Those examples inversely provide some stringent constraints on determining the WF properties.

By taking the RG evolution effects into account, the heavy B meson WF is renormalizable and is an important component for the $k_T$ factorization approach~\cite{libwave}. Under the WW approximation, its transverse and longitudinal momentum dependence is correlated through a simple $\delta$-function, $\delta(\mathbf{k}_\perp^2-\omega(2\bar\Lambda-\omega))$. While by including 3-particle Fock state, the transverse momentum distribution of the B meson WF is broadened to a hyperbola-like curve. The reasonable inclusion of the 3-particle Fock states in B meson WFs provides us with the chance to make a more precise evaluation on the B meson decays. By using the results derived from HQET, we provide a practical framework for constructing the B meson LC WFs $\Psi_{\pm}(\omega,z^2)$. Especially, we present a model in the compact parameter $b$-space. Its behaviors are controlled by two parameters $\bar\Lambda$ and $\delta$. By taking $B\to\pi$ and $B\to K$ TFFs as an example, we show that if the 3-particle WFs' contributions are power suppressed than that of WW case, i.e. its contribution is less than $20\%$, then one obtains preferable values for these two parameters, $\delta\sim 0.27$ and $\bar\Lambda\sim 0.55GeV$. In the literature, the $B\to\pi$ and $B\to K$ TFFs have been studied within the framework of LCSR~\cite{PB2005,bpisr,bpisr2,bpisr3,bpisr4,bpisr5,bpisr6,bpisr7,BDsr4}, a comparison of them with the pQCD estimation can also be adopted for constraining the B meson WF parameters. Further studies on the B meson WFs with higher Fock states and its phenomenological implications are still necessary.

We have suggested a revised light-cone harmonic oscillator model for the WFs of light mesons as $\pi$, $K$, $\eta^{(\prime)}$, $D$ and etc.. By using the constraint from the semi-leptonic process $B \to \pi l \nu$, the parameter $B_\pi$ is restricted to be $[0.01,0.42]$. If taking the process $D \to \pi l \nu$ as a further constraint, we can obtain a more narrow region $B_\pi = [0.00,0.14]$~\cite{TFF7}. For the pion-photon TFF, our present result with the parameter $B_\pi =[0.01,0.14]$ favors the Belle data and the corresponding pion DA has the slight difference from the asymptotic form. Then, one can predict the behavior of the pion-photon TFF in high $Q^2$ regions which can be tested in future experiments.

The $B\to D$ TFF up to twist-4 accuracy by using the improved QCD LCSR with chiral current provides a platform for testing the properties of twist-2 DA. Its second Gegenbauer moment is determined by the parameter $B_D$, $a^D_2 \sim B_D$. By using a proper choice of $B_D$, most of the DA shapes suggested in the literature can be simulated. It is noted that to compare with the experimental result on ${\cal G}(1)\Vcb$, a smaller $B_D \precsim 0.20$ shows a better agreement with the BABAR, CLEO and Belle estimates. By varying $B_D\in[0.00,0.20]$, its first Gegenbauer moment $a^{D}_1$ is about $[0.6,0.7]$. The matrix element $\Vcb$ and its uncertainties have been studied by using two types of processes, e.g. the $B^0/\bar{B}^0$-type and the $B^{\pm}$-type. For the case of $B_D=0$, by adding the errors for all mentioned experimental and theoretical uncertainty sources, we obtain $|V_{cb}| = (40.84\pm3.11)\times 10^{-3}$, where the error stands for the standard derivation of the weighted average. Through a comparison with the experimental data, our present estimation for $\Vcb$ shows a good agreement.

To test the properties of the constructed light pseudoscalar meson DAs, we have discussed the properties of the light pseudoscalar meson-photon TFFs, i.e. $\pi$-$\gamma$, $\eta$-$\gamma$ and $\eta'$-$\gamma$ TFFs. They provide good platforms for studying the DA of the light pseudoscalar mesons. The data from Belle and BABAR have a big difference on the $\pi$-$\gamma$ TFF in high $Q^2$ regions, at present, they are helpless for determining the pion DA. But it is still possible to determine the pion DA as long as we perform a combined analysis of the most existing data of the processes involving pion such as $\pi \to \mu \bar{\nu}$, $\pi^0 \to \gamma \gamma$, $B\to \pi l \nu$, $D \to \pi l \nu$, and etc. More over, we have shown that all pseudoscalar meson-photon transition form factors $Q^{2}F_{\pi\gamma}(Q^2)$, $Q^{2}F_{\eta_q\gamma}(Q^2)$ and $Q^{2}F_{\eta_s\gamma}(Q^2)$ have similar behaviors. No rapid growth has been observed for $Q^{2}F_{\eta\gamma}(Q^2)$ and $Q^{2}F_{\eta'\gamma}(Q^2)$, and these two TFFs can be simultaneously explained by setting $B_{\pi,\eta,\eta'} \sim 0.30$. Possible charm components $f_{\eta}^c$ and  $f_{\eta'}^c$ can shrink the gap between these two TFFs to a certain degree.

As a final remark, in addition to our present LCWF models for the pseudoscalar mesons, such BHL-like model can also be extended to higher twist LCWFs and the LCWFs for other light mesons as the scalars or vectors. In combination with a wide variety of exclusive or semi-exclusive processes involving light mesons, we hope the behaviors of the light mesons' LCWFs and hence their DAs can be determined via global fits of various experimental results finally.

\hspace{1cm}

\noindent{\bf Acknowledgments}: This work was supported in part by Natural Science Foundation of China under Grant No.11075225, No.11275280 and No.11235005, by the Program for New Century Excellent Talents in University under Grant No.NCET-10-0882, and by the Fundamental Research Funds for the Central Universities under Grant No.CQDXWL-2012-Z002.

\end{document}